\documentclass[12pt]{article}
\usepackage{a4wide}
\usepackage{amssymb,amsmath}
\usepackage{graphicx}
\begin{document}
{\renewcommand{\thefootnote}{\fnsymbol{footnote}}
\begin{center}
{\LARGE  Emergent modified gravity: Covariance regained}\\
\vspace{1.5em}
Martin Bojowald\footnote{e-mail address: {\tt bojowald@psu.edu}}
and Erick I.\ Duque\footnote{e-mail address: {\tt eqd5272@psu.edu}}
\\
\vspace{0.5em}
Institute for Gravitation and the Cosmos,\\
The Pennsylvania State
University,\\
104 Davey Lab, University Park, PA 16802, USA\\
\vspace{1.5em}
\end{center}
}

\setcounter{footnote}{0}

\begin{abstract}
  In its canonical formulation, general relativity is subject to gauge
  transformations that are equivalent to space-time coordinate changes of
  general covariance only when the gauge generators, given by the Hamiltonian
  and diffeomorphism constraints, vanish. Since the specific form taken by
  Poisson brackets of the constraints and of the gauge transformations and
  equations of motion they generate is important for general covariance to be
  realized, modifications of the canonical theory, suggested for instance by
  approaches to quantum gravity, are not guaranteed to be compatible with the
  existence of a covariant space-time line element.  This caveat applies even
  if the modification preserves the number of independent gauge
  transformations and the modified constraints remain first class.  Here, a
  complete derivation of covariance conditions, regained from the canonical
  constraints without assuming that space-time has its classical structure, is
  presented and applied in detail to spherically symmetric vacuum models.  As
  a broad application, the presence of structure functions in the constraint
  brackets plays a crucial role, which in an independent analysis has recently
  been shown to lead to higher algebraic structures in hypersurface
  deformations given by an $L_{\infty}$-bracket. The physical analysis of a
  related feature presented here demonstrates that, at least within the
  spherically symmetric setting, new theories of modified gravity are possible
  that are not of higher-curvature form.
\end{abstract}

\section{Introduction}

General relativity cannot be a complete fundamental theory valid on all scales
because a large number of relevant solutions are limited by space-time
singularities. Quantum effects might change this outcome, but they are also
expected to modify general relativity away from singularities. Since general
relativity has a large and non-trivial gauge content, which classically
ensures general covariance, possible modifications that could describe quantum
effects at least in an effective formulation are highly constrained. In a
metric formulation based on space-time tensors, for instance, the class of
admissible effective theories is given by higher-curvature actions. The
observation that the speed of gravitational waves is very close to the speed
of light puts strong constraints on phenomenologically viable higher-curvature
actions \cite{MultiMess1,MultiMess2,MultiMess3,MultiMess4}. It is therefore of
interest to look for new alternatives.

Some aspects of classical gravity and, in particular, of possible
equantizations are more conveniently expressed in a canonical formulation, in
which space-time tensors are replaced by a combination of spatial tensors on
spacelike hypersurfaces in a space-time foliation, with flow equations that
determine how these fields change from hypersurface to nearby
hypersurface. Depending on how they are applied, the flow equations may
present a picture of evolution for the spatial tensors in a given foliation,
or they may be used to determine how the spatial tensors and other quantities
change if one transforms to a different foliation.  For these hypersurface
deformations to be equivalent to general covariance, the spatial tensors on
any hypersurface must obey the Hamiltonian and diffeomorphism constraints of
canonical general relativity \cite{DiracHamGR}.  These constraints, at the
same time, generate the flow equations via their Hamiltonian vector
fields. This equivalence is often used in practice when one interchangeably
refers to coordinate invariance and slicing independence in an analysis of
space-time solutions in general relativity. Both concepts are usually included
in the condition of general covariance, but it turns out that there are subtle
differences between them, owing to the requirement that constraints are to be
imposed.

An immediate implication is that canonical gravity is a Hamiltonian gauge
theory with first-class constraints. However, unlike in gauge theories
encountered for instance in the standard model of particle physics, Poisson
brackets of the constraints do not define a Lie algebra because they do not
have structure constants: In ADM notation \cite{Katz,ADM}, the diffeomorphism
constraint $\vec{H}[\vec{M}]$, depending on a spatial shift vector field
$\vec{M}$ of an infinitesimal tangential deformation of a spatial
hypersurface, and the Hamiltonian constraint $H[N]$, depending on a spatial
lapse function $N$ that determines how much a hypersurface is deformed in its
normal direction within space-time, have Poisson brackets
\begin{eqnarray}
    \{ \vec{H} [ \vec{N}] , \vec{H} [ \vec{M} ] \} &=& \vec{H} [\mathcal{L}_{\vec{N}} \vec{M}]
    \ , \label{eq:Hypersurface deformation or BK algebra in ADM variables}
\\
    \{ H [ N ] , \vec{H} [ \vec{N}]\} &=& - H [ N^b \partial_b N ]
    \ , \\
    \{ H [ N ] , H [ M ] \} &=& \vec{H} [ q^{a b} ( N \partial_b M - M
                                \partial_b N )] \label{HH}
\end{eqnarray}
that depend not only on $\vec{M}$ and $N$, but also on the inverse of the
spatial metric $q_{ab}$ on a spatial hypersurface.

A closed set of brackets is obtained only if the deformation functions
$\vec{M}$ and $N$ inserted in the constraints, including the diffeomorphism
constraint on the right-hand side of (\ref{HH}), depend not only on space-time
coordinates, but independently also on the spatial metric.  Starting with
phase-space independent $\vec{M}$ and $N$ and iterating Poisson brackets, such
as
\begin{eqnarray*}
  &&\{H[N],\{H[N],H[M]\}\}\\
  &=&\{H[N],\vec{H} [ q^{a b} ( N \partial_b M - M
                            \partial_b N )]\}\\
  &=&-H[ q^{a b} ( N \partial_b M - M \partial_b N )\partial_a N]\\
  &&+
      \vec{H}[\{H[N],q^{ab}\}(N\partial_bM-M\partial_bN)]\,, 
\end{eqnarray*}
shows that not only the diffeomorphism constraint in (\ref{HH}) appears with a
metric-dependent shift, but also the Hamiltonian constraint shows up with
a metric-dependent lapse function. Iterating further, different dependencies
on the metric are generated in each step that adds a new factor of the inverse
metric.
However, if we allow phase-space dependent lapse and shift in the Poisson
brackets on the left-hand sides of (\ref{eq:Hypersurface deformation or BK
  algebra in ADM variables})--(\ref{HH}), there are additional terms on the
right-hand sides that, via the chain rule, depend on derivatives of the
deformation functions by the spatial metric. These terms disappear only when
the constraints are imposed and the theory is taken on-shell, giving rise to
the on-shell condition for an equivalence between the gauge symmetries of
hypersurface deformations and coordinate changes. Off-shell, however, there is
a difference between hypersurface deformations and space-time coordinate
transformations.

Mathematically, the dependence on the spatial metric is conveniently expressed
in an algebroid picture, in which equations (\ref{eq:Hypersurface deformation
  or BK algebra in ADM variables})--(\ref{HH}) are related to a suitable
bracket structure for sections of a fiber bundle over the base manifold of
spatial metrics (or a suitable substitute or extension of this space), rather
than a bracket for elements in a Lie algebra as it appears for constraint
brackets without structure functions.  Moreover, as has been shown in an
explicit form only recently \cite{ConsRinehart}, the same metric dependence
also implies that a consistent algebraic bracket corresponding to
(\ref{eq:Hypersurface deformation or BK algebra in ADM variables})--(\ref{HH})
is not Lie but rather an $L_{\infty}$-bracket in which the Jacobi identity is
violated in a specific way. The gauge content of canonical gravity is
therefore described by a higher algebraic structure.

The purpose of the present article, in brief, is to perform a complete
physical analysis of geometrical consequences of structure functions in
hypersurface deformation brackets. As suitable for physical evaluations of
canonical gravity through Hamiltonian vector fields generated by the
constraints, we will employ Poisson brackets, which by definition obey the
Jacobi identity and, when directly applied to the constraint functions on
phase space, do not show the algebraic features of an $L_{\infty}$-bracket. We
will discuss how the structure function of hypersurface deformation brackets
for a given set of modified constraints, such as a general expansion up to a
certain order in derivatives of an effective field theory, can be used to
derive a space-time geometry in which the corresponding constraints generate
hypersurface deformations, and which is subject to a complete set of
covariance conditions. In an analysis of new theories of modified gravity, the
resulting spatial part of the space-time metric may be distinct from the
original phase-space function $q_{ab}$ in which the constraints have been
formulated. Since the precise form of the space-time metric then does not have
a close relationship with the fundamental fields and must be derived using the
form of gauge transformations, it is emergent within this broad set of {\em
  emergent modified gravity}.

This new possibility of modified gravity relies on the presence of structure
functions, just like the higher algebraic structures found earlier. At this
point, however, we are not aware of a more detailed relationship between these
two properties.  From a mathematical point of view, we are looking for
different realizations of the classical algebraic structure underlying
hypersurface deformations, which guarantees that new models will be amenable
to standard space-time analysis using for instance line elements. We are not
interested in modifications of hypersurface deformations or of the underlying
$L_{\infty}$-structure.  Our strategy is comparable to the well-known
derivations of modified gravity in space-time form, which lead to different
realizations of higher-curvature actions that all share the same space-time
structure with standard covariance symmetries, expressed canonically through
hypersurface deformations. The main difference with our approach is that we
aim to derive modified theories fully on the canonical level, arriving at a
space-time picture only at the very end through covariance conditions on the
canonical constraints and their Poisson brackets. Rather surprisingly, we will
show that new modified theories can be obtained in this form that are not of
higher-curvature form. The crucial feature that makes such new theories
possible is that we allow for the resulting (emergent) space-time metric to be
different from the fundamental fields that enter the defining equations, given
here by the constraints. In our case, the correct space-time metric cannot be
identified before a detailed covariance analysis has been performed, in
contrast to other theories of modified gravity in which a space-time metric
must be known before the theory is defined through its curvature tensors. From
a mathematical point of view, the identification of an emergent space-time
makes use of a redefinition of the spatial metric, which could be formulated
as a diffeomorphism on a suitable extension of the base manifold of the
$L_{\infty}$-algebroid, and an application of non-constant sections of the
fiber bundle. These steps in our construction do not change the algebraic
structure of an $L_{\infty}$-algebroid, but they can change the geometry and
physics of space-time solutions of the constraints, equipped with the new
emergent metric.

As a specific example of new theories, brackets of the form
(\ref{eq:Hypersurface deformation or BK algebra in ADM variables})--(\ref{HH})
may be obtained not only for metric-independent $\vec{M}$ and $N$, for which
they have been derived from canonical general relativity via Poisson brackets,
but also for $\vec{M}$ and $N$ with a specific dependence on the spatial
metric and perhaps also extrinsic curvature of a hypersurface. For this to
happen, any contributions from partial derivatives of $\vec{M}$ and $N$ that
initially appear in a calculation of the Poisson brackets would have to cancel
out. There is then a new on-shell interpretation as a gravity theory
associated to these metric and extrinsic-curvature dependent $\vec{M}$ and
$N$, but it need not be the same as the original theory of canonical general
relativity because the structure function, used as the inverse spatial metric
of an emergent space-time line element, need not be of the classical form
where it is identical with one of the basic phase-space degrees of
freedom. This identification of the emergent space-time metric through
structure functions depends on the off-shell behavior of the theory, just as
the higher structures in hypersurface deformations.

In a modified theory of gravity that has a chance of being generally
covariant, the brackets (\ref{eq:Hypersurface deformation or BK algebra in ADM
  variables})--(\ref{HH}) are of the classical form, but possibly with a
modification of the structure function that classically equals inverse spatial
metric $q^{ab}$.  Uniqueness results \cite{Regained,LagrangianRegained} that
show how classical general relativity follows from the brackets
(\ref{eq:Hypersurface deformation or BK algebra in ADM variables})--(\ref{HH})
on-shell can be circumvented by such a modification. Canonical gravity then
has a potential to allow consistent modifications that are not of
higher-curvature form. (All higher-curvature effective actions have the
brackets (\ref{eq:Hypersurface deformation or BK algebra in ADM
  variables})--(\ref{HH}) without a modification of $q^{ab}$
\cite{HigherCurvHam}.) If $q^{ab}$ in (\ref{HH}) is replaced with a different
phase-space function, however, it is not guaranteed that its inverse can still
play the role of a spatial metric in some space-time line element, together
with a lapse function and shift vector for the time components. Our analysis
of modified gravity in canonical form therefore requires an extension of the
classic results of \cite{Regained,LagrangianRegained} in which not only the
dynamical equations, but also space-time structure (that is, the existence of
a consistent space-time line element) must be derived, or regained from the
contraints, their brackets, and from the gauge transformation they
generate. As a general contribution of this paper, we present a complete set
of covariance conditions in canonical form, building on previous constructions
in \cite{EffLine}.

As an example, the Poisson brackets of constraints with phase-space dependent
lapse and shift are guaranteed to equal a linear combination of the
constraints. They are first class and present a consistent gauge theory in
canonical form. For the underlying gauge transformations to correspond to
space-time symmetries via hypersurface deformations, we require in addition
that new contributions to the brackets depending on partial derivatives of
lapse and shift by phase-space degrees of freedom cancel out. A new set of
brackets of the form (\ref{eq:Hypersurface deformation or BK algebra in ADM
  variables})--(\ref{HH}) is then obtained from which a candidate for an
emergent inverse spatial metric can be read off via the structure function.
(We refer to this metric as ``emergent'' in this case because it is not one of
the fundamental fields and must be derived from covariance conditions, unlike
in standard general relativity.)  As we will show, the appearance of a
candidate spatial metric in the brackets does not guarantee that it can be
part of a consistent and coordinate independent space-time line element. We
will derive an additional, previously unrecognized condition on the gauge flow
generated by the Hamiltonian constraint that guarantees matching symmetries
and therefore an invariant emergent space-time line element.  Together with
the cancellation property, this covariance condition imposes strong
restrictions on possible dependences of $\vec{M}$ and $N$ on the spatial
metric or on extrinsic curvature.  We will specify all these conditions and
evaluate them in spherically symmetric models, demonstrating that new theories
of modified gravity are indeed possible in this setting.  Some of the new
models we derive are closely related to recent constructions of consistent
modified theories in canonical spherically symmetric models
\cite{SphSymmMatter,SphSymmMatter2,SphSymmEff,SphSymmEff2}, and they explain
the origin of these modifications.

As a part of our new discussion of general covariance from a canonical theory,
we construct a complete procedure to derive an emergent space-time line
element from canonical hypersurface-deformation brackets, extending previous
results from \cite{EffLine}. In addition to the construction of emergent
modified spherically symmeytric models based on phase-space dependent lapse
and shift, we also construct more general consistent theories that include
potential modifications possibly implied by quantum gravity, such as
non-polynomial terms in extrinsic curvature instead of the classical quadratic
form.  In this case as well, we will see that general covariance imposes
previously unrecognized conditions on possible modifications of canonical
gravity theories, in addition to the usual condition that the constraints
remain first-class and resemble hypersurface-deformation brackets. General
covariance in an emergent line element is therefore recognized as a
restrictive condition on possible quantum space-time effects, required to be
consistent with a geometrical continuum theory of space-time at low
curvature. These general properties will be derived and discussed in
Section~\ref{sec:ModGrav}

We will work out specific versions of generally covariant emergent modified
gravity theories in Section~\ref{sec:SphSymmLin}, using spherically symmetric
reductions. We will first redefine the classical constraints by replacing them
with phase-space dependent linear combinations. The resulting modified
theories demonstrate that off-shell properties of hypersurface deformations
are indeed relevant for physical implications because they make it possible to
evaluate the cancellation condition and the structure function.  We will also
revisit the partial Abelianizations of brackets proposed in
\cite{LoopSchwarz,LoopSchwarz2} in Section~\ref{sec:Partial} and, following
\cite{AbelianStructures,LoopSchwarzCovComm}, re-analyze their off-shell
structure from our new perspective.  Examples of modifications that obey the
new conditions are those of
\cite{LTBII,SphSymmMatter2}. Section~\ref{sec:Modified spherically symmetric
  theory General} will present the general case of modified Hamiltonian
constraints up to second order in spatial derivatives that are consistent with
the covariance condition, including a discussions of the freedom implied by
applying canonical transformations. As an application, in
Section~\ref{sec:PartialGen}, we will use our constraints and related methods
to derive new, non-trivial partial Abelianizations compatible with general
covariance.

\section{Modified gravity in a canonical formulation}
\label{sec:ModGrav}

As usual in canonical theories, we assume that space-time, or at least a
region of interest, is globally hyperbolic: $M = \Sigma \times \mathbb{R}$
with a 3-dimensional ``spatial'' manifold $\Sigma$. In a generally covariant
theory, there is no unique embedding of $\Sigma$ in $M$, but we can
parameterize different choices by working with foliations of $M$ into smooth
families of spacelike hypersurfaces $\Sigma_t$, $t\in{\mathbb R}$, each of
which is homeomorphic to $\Sigma$. For a given foliation, $\Sigma$ can be
embedded in $M$ as a constant-time hypersurface:
$\Sigma\cong \Sigma_{t_0}\cong (\Sigma_{t_0},t_0)\hookrightarrow M$ for any fixed $t_0$.

\subsection{Canonical decomposition}

Given a foliation into spacelike hypersurfaces $\Sigma_t$, a metric
$g_{\mu\nu}$ on $M$ induces a unique spacelike metric $q_{ab}(t_0)$ on any
$\Sigma_{t_0}$. (As in this example, we use Greek letters for indices of
space-time tensors, and Latin letters for indices of spatial tensors.) Using
the unit normal vector field $n^{\mu}$ on $\Sigma_{t_0}$, the induced spatial
metric is obtained by restricting the space-time tensor
$q_{\mu\nu}=g_{\mu\nu}+n_{\mu}n_{\nu}$ to vector fields tangential to the
hypersurface, while $q_{\mu\nu}n^{\nu}=0$. The space-time metric is therefore
expressed as a time-dependent family of spatial metrics. Since spatial
hypersurfaces within a foliation of a covariant theory are invariant under
spatial diffeomorphisms, interpreting the time-dependence of $q_{ab}(t)$ as
unambiguous evolution requires an additional structure that relates points on
infinitesimally related hypersurfaces defined by different values of $t$. This
additional structure can be expressed as a time-evolution vector field
\begin{equation}
  t^\mu = N n^\mu + N^a s_a^\mu 
  \label{eq:Time-evolution vector field}
\end{equation}
in space-time, with the lapse function $N$ and shift vector field $N^a$
\cite{ADM}. The three vector fields $s_a^\mu(t_0)$ inject $T\Sigma_{t_0}$ into
$TM$ such that $g_{\mu \nu} n^\mu s^\mu_a = 0$.

The new ingredients $N$ and $N^a$ describe the frame of an observer in curved
space-time who measures the evolving $q_{ab}(t)$. In the four-dimensional
picture, the frame corresponds to a choice of space-time coordinates which
completes spatial coordinates on $\Sigma$ by a time coordinate $t$ in $M$ such
that $\Sigma_{t_0}=M_{t=t_0}$. The space-time metric or line element is then
in one-to-one correspondence with the family $(q_{ab}(t), N(t), N^a(t))$ of
spatial tensors on the foliation $(\Sigma_t,t)\hookrightarrow M$. We have
\begin{equation}
  {\rm d} s^2 = - N^2 {\rm d} t^2 + q_{a b} ( {\rm d} x^a + N^a {\rm d} t )
  ( {\rm d} x^b + N^b {\rm d} t )
  \,.
  \label{eq:ADM line element}
\end{equation}
A hypersurface in the foliation has extrinsic curvature
\begin{equation} \label{KabLie}
 K_{ab}= \frac{1}{2} {\cal L}_n q_{ab}
\end{equation}
related to the Lie derivative of the spatial metric in the normal
direction. Expressed through a ``velocity'' of $q_{ab}$ with respect to the time-evolution
vector field $t^{\mu}$, it takes the form
\begin{equation}
 K_{ab}  
 = \frac{1}{2N} q_a{}^cq_b{}^d \left({\cal L}_tq_{cd}- {\cal
   L}_Nq_{cd}\right)\,.
\end{equation}

Evolution on a given foliation, defined by a choice of $t$ and $t^{\mu}$ (or
$N$ and $N^a$), is Hamiltonian: Infinitesimal changes of $q_{ab}$ and $K_{ab}$
are obtained via Poisson brackets of these tensors with a Hamiltonian
$H[N,N^a]$, where the Poisson bracket is defined by
considering
\begin{equation}
 p^{ab}=
 \frac{\sqrt{\det q}}{16\pi G} (K^{ab}-K^c_cq^{ab})
\end{equation}
as canonically conjugate momenta of $q_{ab}$. Given the original manifold $M$
as well as general covariance of the relativistic dynamics, evolution within a
foliation is closely related to transformations of the foliation to a new
one. We merely have to re-interpret $N$ and $N^a$ as gauge parameters
$\epsilon^0$ and $\epsilon^a$ that parameterize an infinitesimal change of the
foliation.

Evolution and gauge transformations are therefore described by the
same flow, which implies that the dependence on $q_{ab}$ and $p^{ab}$ of the Hamiltonian
$H[\epsilon^0,\epsilon^a]$ of the gauge flow is the same as the dependence of
the Hamiltonian $H[N,N^a]$ for evolution. In its role as generator of a gauge
flow, however, $H[\epsilon^0,\epsilon^a]$ must be a constraint,
$H[\epsilon^0,\epsilon^a]=0$ for all $\epsilon^0$ and $\epsilon^a$, in order to
have a well-defined symplectic structure on gauge-invariant observables. The
dynamics, therefore, is also fully constrained: $H[N,N^a]=0$ for all $N$ and
$N^a$. (We assume that our manifolds do not require non-trivial boundary
conditions, in which case some choices of $N$ and $N^a$ may not be considered
gauge.) Since spatial and normal deformations of hypersurfaces are
independent, there are two different constraint functionals, the Hamiltonian
constraint $H[N]$ and the diffeomorphism constraint $H_a[N^a]$ such that
$H[N,N^a]=H[N]+H_a[N^a]$. Physical solutions of the theory are ``on-shell,''
that is, they have $q_{ab}$ and $p^{ab}$ on each hypersurface of a foliation
such that $H[N]=0$ and $H_a[N^a]=0$.

The constraints generate gauge transformations for any given phase-space
function $\mathcal{O}$, depending on $q_{ab}$ and $p^{ab}$, via the Poisson
bracket
$\delta_\epsilon \mathcal{O} = \{ \mathcal{O} , H[\epsilon^0 , \epsilon^a]
\}$. These are indeed gauge transformations because the constraints obey the
hypersurface-deformation brackets (\ref{eq:Hypersurface deformation or BK
  algebra in ADM variables})--(\ref{HH}) and are therefore first class. Within
each foliation related by a gauge transformation, the canonical fields
$q_{ab}$ and $p^{ab}$, or any function $\mathcal{O}$ of them, evolve according
to
$\dot{\mathcal{O}} \equiv \delta_t \mathcal{O} = \{ \mathcal{O} , H[N , N^a]
\}$. Poisson brackets of this form do not immediately provide gauge
transformations of $N$ and $N^a$ because they do not have momenta, and they do
not physically evolve because they specify a frame with respect to which
evolution is defined. However, $N$ and $N^a$ must be subject to gauge changes
because the corresponding coefficients in the line element (\ref{eq:ADM line
  element}) depend on the foliation. These gauge transformations can be
derived from the condition that the gauge transformation of an evolution
equation should be consistently related to evolution of gauge-transformed
phase-space variables. This condition refers to commutators of gauge
transformations and evolution, and is therefore sensitive to the structure
functions in (\ref{eq:Hypersurface deformation or BK algebra in ADM
  variables})--(\ref{HH}). Gauge transformations obeying this condition are
given by \cite{PhaseSpaceCoord,LapseGauge,EffLine}
\begin{eqnarray}
    \delta_\epsilon N &=& \dot{\epsilon}^0 + \epsilon^a \partial_a N - N^a \partial_a \epsilon^0
    \ ,\label{eq:Off-shell gauge transformations for lapse and shift}\\
    \delta_\epsilon N^a &=& \dot{\epsilon}^a + \epsilon^b \partial_b N^a - N^b \partial_b \epsilon^a + q^{a b} \left(\epsilon^0 \partial_b N - N \partial_b \epsilon^0 \right)
\end{eqnarray}
where the structure function appears in the last term.

The final ingredient required for a discussion of general covariance in
canonical form is a relationship between gauge transformations generated by a
Hamiltonian and Lie derivatives along a space-time vector field
$\xi^{\mu}$. Components of the latter refer to coordinate directions, while
hypersurface deformations refer to the normal direction. These basis choices
are linearly related by 
\begin{equation}
  \xi^\mu = \epsilon^0 n^\mu + \epsilon^a s^\mu_a
  = \xi^t t^\mu + \xi^a s^\mu_a
  \, ,
\end{equation}
or
\begin{equation}\label{eq:Diffeomorphism generator projection}
  \xi^t = \frac{\epsilon^0}{N}
  \quad ,\quad
  \xi^a = \epsilon^a - \frac{\epsilon^0}{N} N^a\,,
\end{equation}
if we assume that the same spatial coordinate systems are used, as in
(\ref{eq:ADM line element}).  If the constraints and equations of motion are
satisfied (on-shell or ``O.S.''), the gauge transformations
\begin{eqnarray}
    \{ q_{a b} , \vec{H} [\vec{\epsilon}] \} \big|_{{\rm O.S.}}
    = \mathcal{L}_{\vec{\epsilon}} q_{a b}
    \quad,\quad
    \{ q_{a b} , H [\epsilon^0] \} \big|_{{\rm O.S.}}
    = \mathcal{L}_{\epsilon^0 n} q_{a b}
    \label{eq:On-shell gauge transformation of 3-metric}
\end{eqnarray}
together with the gauge transformations of lapse and shift, (\ref{eq:Off-shell
  gauge transformations for lapse and shift}), are equivalent to
infinitesimal space-time diffeomorphisms of the metric in (\ref{eq:ADM line
  element}),
\begin{eqnarray}
    \delta_\epsilon g_{\mu \nu} \big|_{{\rm O.S.}} &=&
    \mathcal{L}_{\xi} g_{\mu \nu}\,,
    \label{eq:Covariance identity - classical GR}
\end{eqnarray}
identifying time derivatives by using evolution equations generated by the
same constraints.

Off-shell, however, hypersurface deformations are rather different from
coordinate changes. The presence of structure functions in the description of
hypersurface deformations implies that a closed set of brackets can be
obtained from them only if lapse and shift are allowed to depend on the
spatial metric, in addition to their dependence on space-time
coordinates. However, if one computes Poisson brackets of the phase-space
functions that provide the Hamiltonian and diffeomorphism constraints smeared
with phase-space dependent lapse and shift, additional terms appear compared
with (\ref{eq:Hypersurface deformation or BK algebra in ADM
  variables})--(\ref{HH}), given by constraints evaluated with partial
derivatives of lapse and shify by components of the spatial metric. These
terms do not change the first-class nature of the constraints or their
on-shell properties, but in general they are not compatible with the
equivalence (\ref{eq:Covariance identity - classical GR}) if one attempts to
extend it to off-shell metrics.  Only specific phase-space functions for lapse
and shift may be compatible with general covariance, provided they obey
conditions that we will derive in what follows. Changing lapse and shift in
this way is equivalent to redefining the Hamiltonian constraint or the normal
direction in a corresponding space-time geometry. The normal direction
together with the spatial metric determines the emergent space-time geometry.
In order to evaluate whether this basic property could lead to new theories of
modified gravity, we have to look look more closely at possible modifications
of the constraints and their resulting brackets.

\subsection{ Hypersurface-deformation brackets and covariance conditions}

There are different sources for possible modifications of the constraints in
models of canonical gravity. As just described, there may be new terms in
their brackets if lapse and shift are allowed to be phase-space dependent.  In
addition, one may be interested in studying possible modifications of the
dependence of the Hamiltonian and diffeomorphism constraints on the canonical
fields.  For instance, higher-order terms beyond the classically at most
quadratic dependence of the constraints on momenta could be motivated by
quantum effects, as a canonical version of higher-curvature effective actions
but without higher time derivatives that usually accompany the latter. We will
derive general conditions for a covariant modification, making two common
assumptions: that the spatial structure of hypersurfaces is unmodified
(governed by the classical diffeomorphism constraint) and that the theory
remains spatially local (with a Hamiltonian constraint that depends on spatial
derivatives up to some finite order).

Any modification of a canonical gauge theory is subject to consistency
conditions. First, the constraints must remain first class, or vanish
on-shell, which is guaranteed if we try to modify the classical theory by
using phase-space dependent lapse and shift, but not necessarily by
modifications of the phase-space dependence of the constraints, as implied by
higher-order terms in momenta.  Secondly, for the modified theory to be
considered a space-time theory, any modified brackets of the constraints must
in some way exhibit an equivalence with space-time coordinate changes, at
least (and usually only) on-shell, as in (\ref{eq:Covariance identity -
  classical GR}). Since the structure functions of classical hypersurface
deformations imply the correct transformations of lapse and shift via
(\ref{eq:Off-shell gauge transformations for lapse and shift}), the modified
brackets must be of the classical form (\ref{eq:Hypersurface deformation or BK
  algebra in ADM variables})--(\ref{HH}).

While this condition leads to brackets identical in form to the classical
ones, an opening for new theories of modified gravity can be found in the
possibility that lapse, shift, and spatial metric as they appear initially in
$H[N]$ and $H[N^a]$ are not required to be identical to the same objects seen
as components of the space-time metric (\ref{eq:ADM line element}), in which
form they have been derived classically. There may be an emergent lapse
$\tilde{N}$, shift $\tilde{N}^a$ and spatial metric $\tilde{q}_{ab}$ that
depend on $N$, $N^a$ and $q_{ab}$ as they appear in the constraints and define
the phase-space structure together with $p^{ab}$, but are not identical to
them. An emergent extrinsic curvature $\tilde{K}_{ab}$ would then be derived
as well for hypersurfaces in the emergent space-time line element defined by
$\tilde{N}$, $\tilde{N}^a$ and $\tilde{q}_{ab}$.  This possibility had been
exploited in \cite{EffLine} to show, via (\ref{eq:Off-shell gauge
  transformations for lapse and shift}), that a sign change of the classical
structure function amounts, under certain conditions, to signature change in
an emergent space-time consistent with general covariance in the modified
theory. The additional conditions, however, were quite restrictive as they
only allowed emergent spatial metrics obtained from $q_{ab}$ by multiplication
with a spatially constant function (which was allowed to depend on time when
used in a completion to a space-time metric as in (\ref{eq:ADM line
  element})).

\subsubsection{Emergent space-time metric and general covariance}

More generally, whenever modifications lead to brackets of the form
\begin{eqnarray}
    \{ \vec{H} [ \vec{N}] , \vec{H} [ \vec{M} ] \} &=& \vec{H} [\mathcal{L}_{\vec{N}} \vec{M}]
    \ , \label{eq:Hypersurface deformation algebra - modified} \\
    \{ \tilde{H} [ N ] , \vec{H} [ \vec{N}]\} &=& - \tilde{H} [ N^b \partial_b N ]
    \ , \\
    \{ \tilde{H} [ N ] , \tilde{H} [ M ] \} &=& - \vec{H} [ \tilde{q}^{a b} (
                                                N \partial_b M - M \partial_b
                                                N )] \label{HHmod}
\end{eqnarray}
without additional terms off-shell for phase-space independent $N$ and
$\vec{M}$, but with a modified structure function $\tilde{q}^{a b}\not=q^{ab}$
as some phase-space function, then $\tilde{q}_{ab}$ rather than $q_{ab}$
should be used as the spatial metric of an emergent space-time line element:
\begin{equation}
    {\rm d} s^2 = - N^2 {\rm d} t^2 + \tilde{q}_{a b} ( {\rm d} x^a + N^a
                    {\rm d} t ) ( {\rm d} x^b + N^b {\rm d} t )
    \,.
    \label{eq:ADM line element - Modified}
\end{equation}
(For now we assume that $\tilde{q}^{ab}$ is invertible; see
Section~\ref{sec:SigChange} for the more general case of $\tilde{q}^{ab}$ that
may be non-invertible on submanifolds of codimension at least one in space-time.)  Under
gauge transformations, lapse and shift then transform as
\begin{eqnarray}
    \delta_\epsilon N &=& \dot{\epsilon}^0 + \epsilon^a \partial_a N - N^a \partial_a \epsilon^0
    \ ,
    \label{eq:Off-shell gauge transformations for lapse - modified}
    \\
    \delta_\epsilon N^a &=& \dot{\epsilon}^a + \epsilon^b \partial_b N^a - N^b \partial_b \epsilon^a + \tilde{q}^{a b} \left(\epsilon^0 \partial_b N - N \partial_b \epsilon^0 \right)
    \ ,
    \label{eq:Off-shell gauge transformations for shift - modified}
\end{eqnarray}
consistent with corresponding coordinate changes in the new emergent line
element.

Here, our construction differs from that in \cite{EffLine}, where factors that
multiply the classical $q^{ab}$ in a modified structure functions were
attempted to be absorbed in a redefined lapse function. Such a choice is more
natural in a discussion of signature change, which is expected to affect the
time components of the space-time metric where the lapse function appears, but
it lead to strong conditions on allowed modifications. Redefining the
spatial metric rather than the lapse function agrees with the constructions of
\cite{SphSymmEff,SphSymmEff2} and earlier in \cite{Absorb}, where signature
change did not occur. Our general treatment here allows for signature change
as well as redefined spatial metrics, as we will see.

A third, and final, condition appears because a modified structure function
$\tilde{q}^{ab}$ is not guaranteed to gauge transform in a way compatible with
an interpretation as the inverse of a spatial metric in a space-time line
element. This condition has not been analyzed completely in previous
studies. We say that the theory is generally covariant if there are
sufficiently many independent fields $f$ (fundamental or composite) such that, (i),
\begin{eqnarray}
    \delta_\epsilon f \big|_{{\rm O.S.}} &=&
    \mathcal{L}_{\xi} f \big|_{{\rm O.S.}}
    \ ,
                                             \label{eq:Covariance condition - general - modified}
\end{eqnarray}
and, (ii), they can be arranged as components of a space-time line element
(\ref{eq:ADM line element - Modified}).  The space-time geometry regained via
(\ref{eq:ADM line element - Modified}) is then generally covariant:
\begin{eqnarray}
    \delta_\epsilon \tilde{g}_{\mu \nu} \big|_{{\rm O.S.}} &=&
    \mathcal{L}_{\xi} \tilde{g}_{\mu \nu} \big|_{{\rm O.S.}}
    \,.
    \label{eq:Covariance condition - modified}
\end{eqnarray}

This covariance condition is not automatically satisfied just by virtue of the
hypersurface deformation brackets, (\ref{eq:Hypersurface deformation algebra -
  modified})--(\ref{HHmod}), even after a redefinition of the spatial metric
or lapse and shift.  In order to see this, we look at each component of the
covariance condition, using (\ref{eq:Hypersurface deformation algebra -
  modified})--(\ref{HHmod}) and performing the ADM decomposition with
(\ref{eq:Diffeomorphism generator projection}).  In what follows, it is
understood that each covariance condition is required to hold only on-shell,
but we drop the symbol ``O.S.''  for the sake of simplicity.

Beginning with the $t a$ components, the left-hand side of the covariance
condition (\ref{eq:Covariance condition - modified}) is
\begin{eqnarray}
    \delta_\epsilon \tilde{g}_{t a}
    = N^b \delta_\epsilon \tilde{q}_{a b}
    + \tilde{q}_{a b} \delta_\epsilon N^b
    \,.
\end{eqnarray}
{\em If} we assume that
$\delta_\epsilon \tilde{q}_{a b} = \mathcal{L}_\xi \tilde{g}_{a b}$, this
covariance condition can be written as
\begin{eqnarray}
    \tilde{q}_{a b} \delta_\epsilon N^b
    &=&
    \mathcal{L}_\xi \tilde{g}_{t a} - N^b \mathcal{L}_\xi \tilde{g}_{a b}
    \nonumber\\
    &=&
    \tilde{q}_{a b} \left( N^b \partial_{t} \xi^t
    + \partial_{t} \xi^b
    - N^b N^c \partial_{c} \xi^t
    - N^c \partial_{c} \xi^b
    + \xi^\mu \partial_\mu N^b \right)
    - N^2 \partial_{a} \xi^t
    \nonumber\\
    &=&
    \tilde{q}_{a b} \left( \dot{\epsilon}^b
    + \epsilon^c \partial_c N^b
    - N^c \partial_{c} \epsilon^b \right)
    + \epsilon^0 \partial_{a} N
    - N \partial_{a} \epsilon^0
    \ ,
\end{eqnarray}
where we have used (\ref{eq:Diffeomorphism generator projection}) in the last
step.  This result is consistent with the canonical gauge transformation of
the shift, (\ref{eq:Off-shell gauge transformations for shift - modified}).
The derivation also shows that the classical relation (\ref{eq:Diffeomorphism
  generator projection}) between a coordinate basis and one adjusted to
hypersurfaces should not be modified: Because there is a term in the final
result for $\tilde{q}_{a b} \delta_\epsilon N^b$ that depends on
$\tilde{q}_{ab}$ and one that does not, all components of this relation have
been used independently in its derivation. (This conclusion might be
circumventd by using metric-dependent coefficients in a modified version of
(\ref{eq:Diffeomorphism generator projection}), but such a choice would
complicate other equations.)

Similarly, the left-hand side of the $t t$ component is
\begin{eqnarray}
    \delta_\epsilon \tilde{g}_{t t}
    = - 2 N \delta_\epsilon N
    + N^a N^b \delta_\epsilon \tilde{q}_{a b}
    + 2 \tilde{q}_{a b} N^a \delta_\epsilon N^b
    \,.
\end{eqnarray}
{\em If} we assume again that $\delta_\epsilon \tilde{q}_{a b} =
\mathcal{L}_\xi \tilde{g}_{a b}$ and that the shift transforms as
(\ref{eq:Off-shell gauge transformations for shift - modified}), then the $t
t$ component of the covariance condition can be written as 
\begin{eqnarray}
    2 N \delta_\epsilon N
    &=&
    2 N \left( \dot{\epsilon}^0
    + \epsilon^a \partial_a N
    - N^a \partial_a \epsilon^0 \right)
    \,,
\end{eqnarray}
which is consistent with the canonical gauge transformation
(\ref{eq:Off-shell gauge transformations for lapse - modified}) of the lapse function.

Lastly, the spatial components of the covariance condition are
\begin{eqnarray}
    \delta_\epsilon \tilde{q}_{a b}
    &=& \mathcal{L}_\xi \tilde{g}_{a b}
    \nonumber\\
     &=&
    \frac{\epsilon^0}{N} \dot{\tilde{q}}_{a b}
    + \epsilon^c \partial_c \tilde{q}_{a b}
    + \tilde{q}_{c a} \partial_{b} \epsilon^c
        + \tilde{q}_{c b} \partial_{a} \epsilon^c\nonumber\\
  &&
    - \frac{\epsilon^0}{N} \left( N^c \partial_c \tilde{q}_{a b}
    + \tilde{q}_{c a} \partial_{b} N^c
    + \tilde{q}_{c b} \partial_{a} N^c \right) \label{deltaq}
\end{eqnarray}
from (\ref{eq:Diffeomorphism generator projection}).
To proceed with our evaluation of this equation, we make the common assumption
that the diffeomorphism constraint remains unmodified, which implies that
$\tilde{q}_{ab}$ is a spatial tensor and that its Poisson bracket with the
diffeomorphism constraint equals a spatial Lie derivative along the shift
vector. The time derivative 
$\dot{\tilde{q}}_{a b} = \{\tilde{q}_{a b} , \tilde{H}[N,N^a]\}$ inserted on the
right-hand side of (\ref{deltaq})  and
the gauge transformation
$\delta_\epsilon \tilde{q}_{a b} = \{\tilde{q}_{a b} ,
\tilde{H}[\epsilon^0,\epsilon^a]\}$ on the left-hand side of this equation then have
matching terms for all spatial derivatives in (\ref{deltaq}) to cancel out. We
are left with the equation
\begin{eqnarray}
    \{\tilde{q}_{a b} , \tilde{H} [\epsilon^0]\} \big|_{\rm O.S.}
    &=& \frac{\epsilon^0}{N} \{\tilde{q}_{a b} , \tilde{H} [N]\} \big|_{\rm O.S.} \,.
    \label{eq:Spatial covariance condition - first reduced form}
\end{eqnarray}

We now assume that the modified theory remains local, such that
$\tilde{H}[\epsilon^0]$ depends on spatial derivatives of the phase-space degrees of
freedom up to some finite order.  As a local functional of $\epsilon_0$, the
normal gauge transformation of the spatial metric takes the generic form
\begin{eqnarray}
    \{\tilde{q}_{a b} , \tilde{H} [\epsilon^0]\} &=&
    Q_{a b} \epsilon^0
    + Q_{a b}^c \partial_c \epsilon^0
    + Q_{a b}^{c d} \partial_c \partial_d \epsilon^0
    + \cdots
    \,,
    \label{eq:Generic normal transformation of 3-metric}
\end{eqnarray}
where the $Q$ tensors are phase-space dependent and the series truncates at
some finite order.
Substituting this expansion into (\ref{eq:Spatial covariance condition - first
  reduced form}), we obtain
\begin{eqnarray}
&&    Q_{a b}^c \frac{\partial_c \epsilon^0}{\epsilon^0}
    + Q_{a b}^{c d} \frac{\partial_c \partial_d \epsilon^0}{\epsilon^0}
    + \cdots \bigg|_{{\rm O.S.}}\nonumber\\
    &=&
    Q_{a b}^c \frac{\partial_c N}{N}
    + Q_{a b}^{c d} \frac{\partial_c \partial_d N}{N}
    + \cdots \bigg|_{{\rm O.S.}}
    \label{eq:Spatial covariance condition - second reduced form}
\end{eqnarray}
(neglecting boundary terms that may result after integrating by parts).  For a
generally covariant theory, the gauge generator functions
$(\epsilon^0,\epsilon^a)$ must be independent of each other, and of $N$, $N^a$
as well as phase-space functions that they are supposed to transform.  Thus,
each $Q$ tensor in (\ref{eq:Generic normal transformation of 3-metric}) must
vanish independently, and we obtain a series of conditions on the gauge
transformation of the emergent spatial metric or its inverse:
\begin{eqnarray}
    \frac{\partial (\delta_{\epsilon^0} \tilde{q}^{a b})}{\partial (\partial_c
  \epsilon^0)} \bigg|_{{\rm O.S.}}
    = \frac{\partial (\delta_{\epsilon^0} \tilde{q}^{a b})}{\partial
  (\partial_c \partial_d \epsilon^0)} \bigg|_{{\rm O.S.}}
    = \cdots
    = 0
    \,.
    \label{eq:Covariance condition - modified - reduced}
\end{eqnarray}
Since this tensor is determined by the structure function in the
hypersurface-deformation brackets of a modified canonical theory, generally
covariant modifications are subject to additional conditions that go beyond the
basic requirement that the brackets remain first class. They must be first
class with structure functions obeying (\ref{eq:Covariance condition - modified - reduced}).
Because the structure function and its gauge transformation are both
determined by the constraints, these are non-trivial conditions on the
modified Hamiltonian constraint.  This condition has been overlooked in
several previous treatments of canonical gravity and possible modifications,
such as models of loop quantum gravity, but it is essential for complete
covariance.

\subsubsection{Extrinsic curvature and the geometry of embedded hypersurfaces}

In our discussion so far, we have used the constraint equations and the gauge
flow (\ref{eq:Spatial covariance condition - first reduced form}) or the
equation of motion of the emergent metric. The condition that this
transformation is equivalent to the Lie derivative of a spatial metric led to
a non-trivial consistency condition (\ref{eq:Covariance condition - modified -
  reduced}) for general covariance. Owing to modifications of the structure
function, the emergent spatial metric that obeys this condition need not agree
with the basic phase-space function $q_{ab}$ of the canonical theory, unlike
in classical theories of gravity. In the same way, the momentum $p^{ab}$
canonically conjugate to $q_{ab}$ need not be linearly related to extrinsic
curvature if the structure function is modified. Instead, the emergent
space-time line element (\ref{eq:ADM line element - Modified}) may be used to
derive a suitable extrinsic curvature tensor $\tilde{K}_{ab}$ in the same
slicing in which the emergent metric $\tilde{q}_{ab}$ is induced by
(\ref{eq:ADM line element - Modified}) as the spatial metric. Since extrinsic
curvature by definition depends on normal derivatives of the spatial metric,
equations of motion of the canonical theory could be used to relate
$\tilde{K}_{ab}$ to the original canonical fields $q_{ab}$ and $p^{ab}$, just
as $\tilde{q}_{ab}$ is not a basic canonical field but depends, in general, on
both $q_{ab}$ and $p^{ab}$ in a non-linear way.

Such an extrinsic curvature tensor $\tilde{K}_{ab}$ would be derived from a
consistent space-time geometry and would therefore be a proper co-variant
2-tensor. This property implies that there is no additional momentum-version
of the covariance condition derived from $\tilde{q}_{ab}$. If one is
interested in comparing canonical gauge transformations of $\tilde{K}_{ab}$,
defined through the relationship between this tensor and the phase-space
functions $q_{ab}$ and $p^{ab}$, with space-time coordinate transformations of
this tensor, one would make use of the momentum version of the gauge
transformations (or of the corresponding equations of motion) (\ref{eq:Spatial
  covariance condition - first reduced form}). The full set of gauge
transformations is therefore used if one compares gauge transformations with
space-time Lie derivatives for both $\tilde{q}_{ab}$ and
$\tilde{K}_{ab}$. However, since the structure function of
hypersurface-deformation brackets uniquely determines the complete space-time
line element, only the gauge transformations of $\tilde{q}_{ab}$ yield
non-trivial covariance conditions, while covariance of $\tilde{K}_{ab}$ is
then implied. Heuristically, the correct transformation of $\tilde{K}_{ab}$ is
implied because $\tilde{K}_{ab}$ is defined as a certain space-time coordinate
change of $\tilde{q}_{ab}$. (This transformation is not a Lie derivative
because extrinsic curvature depends on the slicing. It is a spatial tensor on
a fixed hypersurface but not a space-time tensor.)  A detailed derivation
together with explicit equations for  the correct coordinate transformations can be
found in App.~\ref{a:ExtCurv}.

Covariance of the spatial metric tensor therefore implies covariance of the
extrinsic-curvature tensor. While all equations of motion are used if one
derives explicit transformations for both $\tilde{q}_{ab}$ and
$\tilde{K}_{ab}$, only the former lead to non-trivial covariance
conditions. This result re-inforces the heuristic understanding that the
equations of motion for $\tilde{q}_{ab}$ determine geometrical properties of
an embedded hypersurface (the relationship between extrinsic curvature and
normal derivatives of the spatial metric), equations of motion for
$\tilde{q}_{ab}$ determine the dynamics of the theory and therefore physical
properties. 

\subsubsection{A simple example}

As an example, consider a theory in metric variables, where the phase space is
composed of the ``bare'' spatial metric $q_{a b}$ (used to define the
phase-space structure) and its conjugate momenta $p^{a b}$, and the emergent
spatial metric equals the bare spatial metric (that is, the structure function
remains classical).  The covariance condition (\ref{eq:Covariance condition -
  modified - reduced}) then implies, from
$\{q_{a b} , \tilde{H}[\epsilon^0]\} = \delta \tilde{H}[\epsilon^0] / \delta
p^{a b}$, that the Hamiltonian constraint must not contain spatial derivatives
of $p^{a b}$.  If we use only up to second-order spatial derivatives of
$q_{ab}$, the Hamiltonian constraint is uniquely determined by the
hypersurface deformation brackets, (\ref{eq:Hypersurface deformation algebra -
  modified})--(\ref{HHmod}), up to the choice of Newton's and the cosmological
constant, and assuming parity symmetry \cite{Regained,LagrangianRegained}. It
must therefore be classical, and generally covariant modifications are ruled
out under the stated conditions.

If the spatial metric is considered a composite function of the phase space,
as it happens when the space-time line element has an emergent spatial metric
$\tilde{q}_{ab}$ distinct from the phase-space function $q_{ab}$ (and not just
obtained by directly applying a canonical transformation), the regaining
procedure of \cite{Regained,LagrangianRegained} is modified and may result in
new gravitational theories even at second derivative order. The covariance
condition (\ref{eq:Covariance condition - modified - reduced}) is non-trivial
in this situation.  For instance, if the emergent spatial metric depends on
the momenta, as it happens in the examples discussed in the next section,
spatial derivatives of the bare metric, which always appear in the Hamiltonian
constraint, also contribute to the covariance condition and could cancel out
with unwanted terms from spatial derivatives of the momenta.  The covariance
condition and the concept of an emergent metric then present important
ingredients in constructions of modified canonical gravity, in addition to the
requirement that hypersurface-deformation brackets of the form
(\ref{eq:Hypersurface deformation algebra - modified}) be realized. Although
there are then different versions of $q_{ab}$ in such a theory, given by the
bare metric and the emergent metric, it is not an example of bi-metric
gravity: A unique metric, the emergent one, is singled out by the covariance
condition, if the latter can be solved at all.

Results similar to those of the present section can be formulated for triad
variables, in which case the spatial metric has the status of a composite
field even in the classical theory.

\subsection{Non-invertible structure functions and signature change}
\label{sec:SigChange}

The definition of an emergent line element from a modified structure function
$\tilde{q}^{ab}$ requires that this spatial tensor is invertible in a
space-time region in which it is applied. If this condition is not strictly
fulfilled but still holds on a dense submanifold in space-time, there are
hypersurfaces (possibly timelike or lightlike and not just spacelike) that
separate regions in which $\tilde{q}^{ab}$ is invertible.  Emergent line
elements then exist only in these regions but not on the separating
hypersurfaces. Moreover, they may differ from one another by certain sign
factors of ${\rm sgn}\det\tilde{q}^{ab}$. Space-time then has distinct regions
in which emergent line elements exist, but no global line element.  (A single
space-time object is defined by solutions of the constraints before they are
equipped with emergent line elements.)  An example for such emergent
space-times is given by models with dynamical signature change
\cite{Action,SigChange}.

If we are in a region where ${\rm sgn}\det\tilde{q}^{ab}=-1$, the emergent
line element (\ref{eq:ADM line element - Modified}) is of (negative) Euclidean
signature $(-1,-1,-1,-1)$ and no longer Lorentzian as in the classical
limit. (We assume that the signature remains spatially isotropic in order to
prevent the existence of a distinguished spatial direction in the resulting
gravity theory.)  Combining our constructions with those in \cite{EffLine}, it
follows that this emergent line element is equivalent to one with positive
Euclidean signature $(+1,+1,+1,+1)$ because we may define the emergent spatial
metric as
\begin{equation}
  \tilde{\tilde{q}}_{ab}= {\rm sgn}(\det\tilde{q}^{ab})\tilde{q}_{ab}
\end{equation}
and introduce an emergent line element
\begin{equation}
    {\rm d} s^2 = -{\rm sgn}(\det\tilde{q}^{ab}) N^2 {\rm d} t^2 +
    \tilde{\tilde{q}}_{a b} ( {\rm d} x^a + N^a 
                    {\rm d} t ) ( {\rm d} x^b + N^b {\rm d} t )
    \,.
\end{equation}
Because ${\rm sgn}(\det\tilde{q}^{ab})$ is spatially constant in any region in
which $\tilde{q}^{ab}$ is invertible, the conclusions of \cite{EffLine} apply
and show that the new definitions guarantee general covariance.

\section{Spherically symmetric theory of gravity in vacuum}
\label{sec:SphSymmLin}

Following the general results of the previous section, we will now focus on
spherically symmetric models. We choose the basic phase-space variables
to be certain extrinsic-curvature components as the configuration variables
and densitized-triad components as their conjugate momenta, as used frequently
in models of loop quantum gravity \cite{SphSymm,SphSymmHam}.

\subsection{Classical spherically symmetric theory}

In the spherically symmetric classical theory, the spacetime metric is
\begin{equation}
    {\rm d} s^2 = - N^2 {\rm d} t^2 + q_{x x} ( {\rm d} x + N^r {\rm d} t
                    )^2 + q_{\vartheta \vartheta} {\rm d} \Omega^2 
\end{equation}
where ${\rm d}\Omega^2={\rm d}\vartheta^2+ \sin^2\vartheta {\rm d}\varphi^2$.
The spatial metric components are related to the radial and angular components
of a densitized triad, $E^x$ and $E^{\varphi}$, respectively, via
\begin{equation}
  q_{x x} = \frac{(E^\varphi)^2}{E^x} \quad,\quad
    q_{\vartheta \vartheta} = E^x
    \ ,\label{eq:ADM metric - spherical}
\end{equation}
which, for the purpose of the present paper, may be considered a
part of a canonical transformation of the phase space in metric
variables. Extrinsic-curvature components are then transformed to radial
fields $K_x$ and $K_{\varphi}$ such that
\begin{equation}
    \{ K_x (x) , E^x (y)\}
    = \{ K_\varphi(x) , E^\varphi (y) \}
    = \delta (x-y) 
    \,.
  \end{equation}
  (We choose units such that $2G=1$.)
The geometrical interpretation of $K_x$ and $K_{\varphi}$ follows from their
equations of motion, generated by the Hamiltonian and diffeomorphism constraints,
\begin{eqnarray}
    H [N] &=& \int {\rm d} x\ N \bigg[ \frac{((E^x)')^2}{8 \sqrt{|E^x|} E^\varphi}
    - \frac{E^\varphi}{2 \sqrt{|E^x|}}
    - \frac{E^\varphi K_\varphi^2}{2 \sqrt{|E^x|}}
    - 2 K_\varphi \sqrt{|E^x|} K_x 
    \nonumber\\
    &&\qquad - \frac{\sqrt{|E^x|} (E^x)' (E^\varphi)'}{2 (E^\varphi)^2} 
    + \frac{\sqrt{|E^x|} (E^x)''}{2 E^\varphi}
    \bigg] 
    \, ,
    \label{eq:Hamiltonian constraint - spherical symmetry - Gauss solved}
\end{eqnarray}
and
\begin{eqnarray}
    H_r [N^r] &=& \int {\rm d} x\ N^r \left( K_\varphi' E^\varphi - K_x (E^x)' \right)
    \, ,
    \label{eq:Diffeomorphism constraint - spherical symmetry- Gauss solved}
\end{eqnarray}
where the primes are radial derivatives.

The hypersurface-deformation brackets in this case are
\begin{eqnarray}
    \{ H_r [N^r] , H_r[M^r] \}&=& H_r [N^r (M^r)' - M^r (N^r)']
    \ , \label{eq:Hypersurface deformation algebra - spherical}
\\
    \{ H [N] , H_r [M^r] \}&=& - H[M^r N'] 
    \ ,
    \label{eq:H,H_r bracket}
    \\
    \{ H [N] , H[M] \}&=& H_r \left[ q^{x x} \left( N M' - N' M \right)\right]
    \label{eq:H,H bracket}
\end{eqnarray}
with the structure function $q^{x x} = E^x/(E^\varphi)^2$, which indeed
follows from Poisson brackets of the constraints for phase-space independent
$N$ and $M^r$. The structure function determines
off-shell gauge transformations for lapse and shift as
\begin{eqnarray}
    \delta_\epsilon N &=& \dot{\epsilon}^0 + \epsilon^r N' - N^r (\epsilon^0)' ,\\
    \delta_\epsilon N^r &=& \dot{\epsilon}^r + \epsilon^r (N^r)' - N^r (\epsilon^r)' + q^{x x} \left(\epsilon^0 N' - N (\epsilon^0)' \right) .\label{eq:Off-shell gauge transformations for lapse and shift - spherical}
\end{eqnarray}

Condition (\ref{eq:Covariance identity - classical GR}) for the covariance of
the metric is satisfied, and the gauge generator functions are related to the
2-component vector generator of infinitesimal diffeomorphisms by
\begin{equation}
    \xi^\mu = \epsilon^0 n^\mu + \epsilon^r s^\mu = \xi^t t^\mu + \xi^r s^\mu
\end{equation}
with components
\begin{equation}
    \xi^t = \frac{\epsilon^0}{N}
    \quad , \quad
    \xi^r = \epsilon^r - \frac{\epsilon^0}{N} N^r
    \ .
    \label{eq:Diffeomorphism generator projection - spherical}
\end{equation}

\subsection{Modified spherically symmetric theory}
\label{sec:Modified spherically symmetric theory}

We consider modifications to the spherically symmetric theory with canonical
variables $(K_\varphi , E^\varphi)$ and $(K_x , E^x)$.  If we modify the
Hamiltonian constraint, the brackets (\ref{eq:Hypersurface deformation algebra
  - spherical})--(\ref{eq:H,H bracket}) determine the emergent radial spatial
metric via $\tilde{q}_{x x} = (\tilde{q}^{x x})^{-1}$ if the modified
structure function $\tilde{q}^{xx}$ is invertible everywhere, and therefore
positive definite. The angular component of the metric
$\tilde{q}_{\vartheta \vartheta}$ cannot be determined in this way because it
does not appear in the classical brackets. We will therefore keep it
unmodified in the present section.  The emergent space-time line element is
then given by
\begin{eqnarray}\label{eq:ADM line element - spherical - modified}
    {\rm d} s^2 &=& - N^2 {\rm d} t^2 + \tilde{q}_{x x} ( {\rm d} x + N^r {\rm
                    d} t )^2 + E^x {\rm d} \Omega^2
\end{eqnarray}
if $\tilde{q}^{xx}>0$ is strictly positive. If $\tilde{q}^{xx}$ is not
positive definite, we define the emergent line element as
\begin{eqnarray}
    {\rm d} s^2 &=& -{\rm sgn}(\tilde{q}^{xx}) N^2 {\rm d} t^2 + \tilde{\tilde{q}}_{x
                    x} ( {\rm d} x + N^r {\rm 
                    d} t )^2 + E^x {\rm d} \Omega^2
\end{eqnarray}
with $\tilde{\tilde{q}}_{xx}=|\tilde{q}^{xx}|^{-1}$, choosing the second
option of Section~\ref{sec:SigChange} in order to avoid a distinguished role
played by the radial direction in space-time signature. This choice is
determined by our decision to keep $q_{\vartheta\vartheta}$ unmodified.

Another immediate implication of this decision is that
$q_{\vartheta\vartheta}=E^x$ is not a composite field in the emergent line
element. As in our general discussion, we therefore conclude that modified
constraints cannot depend on spatial derivatives of the variable $K_x$
canonically conjugate to $E^x$: The covariance condition (\ref{eq:Covariance
  condition - modified - reduced}), evaluated for the angular component of the
metric, implies
\begin{eqnarray}
    \frac{\partial \tilde{H}}{\partial K_x'} \bigg|_{{\rm O.S.}}
    = \frac{\partial \tilde{H}}{\partial K_x''} \bigg|_{{\rm O.S.}}
    = \cdots
    = 0
    \ ,
    \label{eq:Covariance condition on K_x - modified - spherical}
\end{eqnarray}
using $\delta_{\epsilon^0} E^x = - \delta \tilde{H} [\epsilon^0]/\delta K_x$.

Radial derivatives of $K_x$ in $\tilde{H}$ can be consistent with the
covariance condition only if one considers a more general emergent angular
metric component $\tilde{q}_{\vartheta \vartheta}$ that depends not only on
$E^x$ but also on other phase-space variables such as extrinsic
curvature. Within spherically symmetric models, a choice of
$\tilde{q}_{\vartheta \vartheta}\not=E^x$ could therefore be justified if one
would like to include a specific term, for instance with spatial derivatives
of $K_x$, in a modified Hamiltonian constraint. Such terms have been
considered in \cite{HigherSpatial} but without finding a closed version of the
modified constraints. Alternatively, a modified angular metric could
potentially be determined by constraint brackets if they are derived from the
spherical reduction of a consistently modified full theory, or from a model
system with less symmetry than spherical models.  We will leave these
possibilities for future research.

The radial component of the covariance condition takes the form
\begin{eqnarray}
    \frac{\partial (\delta_{\epsilon^0} \tilde{q}^{x x})}{\partial
  (\epsilon^0)'} \bigg|_{{\rm O.S.}}
    = \frac{\partial (\delta_{\epsilon^0} \tilde{q}^{x x})}{\partial
  (\epsilon^0)''} \bigg|_{{\rm O.S.}}
    = \cdots
    = 0
    \,.
    \label{eq:Covariance condition - modified - spherical}
\end{eqnarray}
Since $\tilde{q}^{xx}$, like $q^{xx}$ itself in a triad formulation, is a
composite field, this condition is more complicated than the angular version
(\ref{eq:Covariance condition on K_x - modified - spherical}).  We will
consider specific modified constraints and their structure functions in our
evaluations of this condition.

A direct application of the covariance condition (\ref{eq:Covariance condition
  - modified - spherical}) to the emergent space-times considered in
\cite{LTBII} and \cite{SphSymmMatter2} confirms that these two models are both
covariant.  The space-times proposed in several other works, among them
\cite{LoopSchwarz2,EffLine}, can be shown not to satisfy covariance, even
though the underlying modified constraints are first class and have constraint
brackets of hypersurface-deformation form.  In the next subsection we
construct a new example by using phase-space dependent gauge generator
functions, as performed in a different way in \cite{LoopSchwarz,LoopSchwarz2}.

\subsection{Linear combination of constraints with phase-space dependence}
\label{sec:Lin}

We have now specified conditions for general covariance of an emergent line
element determined by the structure functions of hypersurface-deformation
brackets.  As a first application, we can now test whether it is possible, at
least in spherically symmetric models, to construct modified gravity theories
by using different versions of phase-space dependent lapse and shift in such a
way that the structure function no longer agrees with a basic phase-space
variable.

From the point of view of a canonical gravity theory, phase-space dependent
lapse and shift matter because they imply additional terms in off-shell gauge
transformations.  Consider two phase space functions $Q$ and $B$, and a lapse
function $N$ that is phase-space independent. The gauge transformation of $Q$
generated by the Hamiltonian constraint $H$ with gauge function $B N$ instead
of $N$ is given by
\begin{eqnarray} \label{QBN}
    &&\{ Q , H [B N] \} = \int {\rm d} y \ \{ Q , H(y) B (y) \}
                          N(y)\\
    &=& \int {\rm d} y \ \big( \{ Q , H(y) \}  B (y) + \{ Q , B (y) \} H(y)
        \big) N(y) \nonumber
\end{eqnarray}
with a new term $\{Q,B\} \neq 0$. While this new term is multiplied by $H(y)$
and therefore disappears on-shell, it changes the form of off-shell gauge
transformations. Off-shell gauge transformations, applied to the constraints
themselves, are relevant for properties of hypersurface-deformation brackets
and may contribute to their structure functions and thereby to emergent line
elements. Our new methods from the previous section bring us in a position to
evaluate these implications.

For the sake of simplicity, we will implement phase-space dependent lapse and shift in a
way that does not change spatial diffeomorphisms. Only the Hamiltonian
constraint will then have a phase-space dependent multiplier. As a generalization of
(\ref{QBN}), in addition to replacing $N$ with $BN$ we may also add a
contribution from the diffeomorphism constraint to the new normal
deformation. Formally, we may arrive at such linear combinations by the
substitution
\begin{equation} \label{subs}
  N\to BN \quad,\quad N^r\to AN+N^r
\end{equation}
in the original constraints with phase-space independent $N$ and $N^r$. The
original constraints obey the brackets (\ref{eq:Hypersurface deformation
  algebra - spherical}), while phase-space dependent $A$ and $B$ imply
additional terms in the brackets of $H_r[N^r]$ with a new Hamiltonian
constraint
\begin{equation} \label{Hnew}
    H^{{\rm (new)}} [N] = H[B N] + H_r [A N]
\end{equation}
derived from the complete Hamiltonian
\begin{eqnarray}
    H[B N , A N + N^r] = H [B N] + H_r [A N] + H_r [N^r]
     \label{eq:Total Hamiltonian - Linear combination}
\end{eqnarray}
after the substitution (\ref{subs}), collecting all $N$-dependent terms in the
definition of $H^{({\rm new})}[N]$. This procedure of implementing a
phase-space dependent linear combination of the constraints follows a
construction proposed in \cite{LoopSchwarz,LoopSchwarz2} that allows one to
eliminate structure functions. Off-shell consistency conditions and
covariance, however, had not been considered in these papers, which is made
more difficult, if not impossbile, by the very act of eliminating the
structure function that determines consistent space-time structures. In the
present paper we are not interested specifically in eliminating structure
functions, but our methods are general enough to analyze covariance also in
this context.  We will briefly return to this question after our derivation of
general consequences of phase-space dependent linear combinations.

In general, the off-shell Poisson brackets of $H^{({\rm new})}[N]$ and
$H_r[N^r]$ are not of the form (\ref{eq:Hypersurface deformation algebra -
  spherical}).  The existence of a covariant emergent line element is
therefore not guaranteed. We will now use our methods from the previous
section to derive new results that tell us under which conditions on $A$ and
$B$ a covariant emergent line element exists, based on (\ref{eq:Covariance
  condition on K_x - modified - spherical}) and (\ref{eq:Covariance condition
  - modified - spherical}).

\subsubsection{Anomaly-freedom and the covariance condition}

We now consider the same canonical variables $(K_\varphi , E^\varphi)$ and
$(K_x , E^x)$ as well as the diffeomorphism constraint $H_r$ from
(\ref{eq:Diffeomorphism constraint - spherical symmetry- Gauss solved}) as
used in spherically symmetric gravity. Our derivations in this subsection are
general enough to allow for a generic initial Hamiltonian constraint
$H^{{\rm (old)}}$ that could, for instance, correspond to a dilaton gravity
model. We will then implement a phase-space dependent linear transformation of
the form just described, replacing $H^{({\rm old)}}$ with $H^{({\rm new})}$
defined as in (\ref{Hnew}).  Because $H^{{\rm (old)}}$ and $H^{{\rm (new)}}$
(before smearing) are densities of weight one and $H_r$ is a density of weight
two, it follows that $B$ has density weight zero and that $A$ has density
weight $-1$.

By construction, the gauge transformations $\delta_\epsilon^{{\rm (new)}}$
generated by the new constraint (\ref{Hnew}) are equivalent to a combination
of gauge transformations generated by the old constraint and the
diffeomorphism constraint, with partially phase-space dependent generators:
\begin{eqnarray}
    \delta_{\epsilon}^{{\rm (new)}} \equiv \delta_{\epsilon^0 ,
  \epsilon^r}^{{\rm (new)}} = \delta^{{\rm (old)}}_{B \epsilon^0 , A \epsilon^0 +  \epsilon^r } \ .
\end{eqnarray}
In order to highlight new terms implied by phase-space dependent multipliers,
we define for label=``new'' and label=``old'' the contribution
\begin{eqnarray}
    \not{\delta}_{F_0 \epsilon^0 , F_r \epsilon^r}^{{\rm (label)}} Q
  &\equiv& \int {\rm d} y \ \{ Q , H^{{\rm (label)}} (y) \} F_0 (y) \epsilon^0 (y)
    + \int {\rm d} y \ \{ Q , H_r (y) \} F_r (y) \epsilon^0 (y)
\end{eqnarray}
to normal gauge transformations, where $\epsilon^0$ and $\epsilon^r$ are
phase-space independent, while $Q$, $F_0$, and $F_r$ are phase-space
functions.  Assuming that we already know the gauge transformations generated
by the old constraints, we can then write the new transformations as
\begin{eqnarray}
    \delta_\epsilon^{{\rm (new)}} Q (x)
    &=& \not{\delta}^{{\rm (old)}}_{B \epsilon^0 , A \epsilon^0 + \epsilon^r} Q (x)
    + H^{{\rm (old)}} \left[ \{ Q (x) , B \} \epsilon^0  + \{ Q (x) , A \} \epsilon^r \right]
    + H_r^{{\rm (old)}} \left[ \{ Q (x) , A \} \epsilon^0 \right]\,.
\end{eqnarray}

We begin with the condition that the bracket (\ref{eq:H,H_r bracket}) should
be re-obtained for $H^{{\rm (new)}}$ and $H_r$ if the new constraints
correspond to a realization of hypersurface deformations.
A direct calculation shows
\begin{eqnarray}
    \{ H^{{\rm (new)}} [N] , H_r [M^r] \} &=&
    - H^{{\rm (new)}} \left[ M^r N' \right] 
    \nonumber\\
    &&+ H_r \left[
    ( \delta_{0 , M^r}^{{\rm (old)}} A ) N - \left( A' M^r - A (M^r)' \right) N 
    \right]
    \nonumber\\
    &&+ H^{{\rm (old)}} \left[ ( \delta_{0 , M^r}^{{\rm (old)}} B ) N - B' M^r N \right]
    \ .
\end{eqnarray}
For this to be of the hypersurface deformation form, we have the conditions
\begin{eqnarray}\label{eq:H,H_r bracket conditions}
    \delta_{0 , M^r}^{{\rm (old)}} A &=& M^r A' - A (M^r)'
    \ ,\\
    \delta_{0 , M^r}^{{\rm (old)}} B &=& B' M^r \label{BHr} \ .
\end{eqnarray}
which are satisfied provided $B$ is of density weight zero and $A$ of
density weight $-1$.

The bracket of the new Hamiltonian constraint with itself contains several terms:
\begin{eqnarray}
&&    \{ H^{{\rm (new)}} [N] , H^{{\rm (new)}} [M] \}\\
    &=&
    \int {\rm d} x {\rm d} y \ N (x) M (y) \Bigg(
    \left\{ H^{({\rm old})} (x) , H^{({\rm old})} (y) \right\} B(x) B(y)
    + \left\{ B(x) , B(y) \right\} H^{({\rm old})} (x) H^{({\rm old})} (y)
    \nonumber\\
    &&+ \bigg( \left\{ B(x) , H^{({\rm old})} (y) \right\} H^{({\rm old})} (x) B(y) - (N \leftrightarrow M) \bigg)
    \nonumber\\
    &&+ \left\{ H_r (x) , H_r (y) \right\} A(x) A(y)
    + \left\{ A(x) , A(y) \right\} H_r (x) H_r (y)
    \nonumber\\
    &&+ \bigg( \left\{ A (x) , H_r (y) \right\} H_r (x) A(y) - (N \leftrightarrow M) \bigg)
    \nonumber\\
    &&+ \bigg( \left\{ H_r (x) , H^{({\rm old})} (y) \right\} A(x) B(y)
    + \left\{ A(x) , H^{({\rm old})} (y) \right\} H_r (x) B(y)
    - (N \leftrightarrow M) \bigg)
    \nonumber\\
    &&+ \bigg( \left\{ B (x) , H_r (y) \right\} H (x) A (y)
    + \left\{ A(x) , B(y) \right\} H_r (x) H^{({\rm old})} (y)
    - (N \leftrightarrow M) \bigg)
    \Bigg)\,.
  \nonumber
\end{eqnarray}
The first term contains the old bracket
$\{H^{({\rm old})}(x),H^{({\rm old})}(y)\}$ for which we can use the
hypersurface-deformation result, but there are several additional terms which
can be written as
\begin{eqnarray}
 \{ H^{{\rm (new)}} [N] , H^{{\rm (new)}} [M] \}
    &=&
    H_r \left[ B^2 q^{x x}_{{\rm (old)}} \left( N M' - M N' \right) 
    + \left(\not{\delta}_{B M,0}^{{\rm (old)}} A \right) N -
        \left(\not{\delta}_{B N,0}^{{\rm (old)}} A \right) M \right]
    \nonumber\\
    &&+ \int {\rm d} x {\rm d} y \ \bigg( \left\{ B (x) , A (y) \right\} N (x)
       H^{{\rm (old)}} (x) M (y) H_r (y) 
    - (N \leftrightarrow M) \bigg)
    \nonumber\\
    &&+ \int {\rm d} x {\rm d} y \ \left\{ A(x) , A(y) \right\} N (x) H_r (x) M (y) H_r (y)
    \nonumber\\
    &&+ \int {\rm d} x {\rm d} y \
    \left\{ B(x) , B(y) \right\} N (x) H^{{\rm (old)}} (x) M (y) H^{{\rm (old)}} (y)
    \nonumber\\
    &&+ H^{{\rm (old)}} \left[ \left(\not{\delta}_{B M,0}^{{\rm (old)}} B \right) N
    - \left(\not{\delta}_{B N,0}^{{\rm (old)}} B \right) M
    + A B \left( N M' - M N' \right) \right]
    \,.
    \label{eq:H,H bracket of new constraints}
\end{eqnarray}
For this combination of terms to be of the required hypersurface-deformation
form, it must include $H_r$ smeared by $q^{x x}_{{\rm (new)}} (N M' - M N')$
where $q^{x x}_{{\rm (new)}}=\tilde{q}^{xx}$ is the new structure function on
which we will impose our condition for general covariance. These terms are
contained in the first three lines of (\ref{eq:H,H bracket of new
  constraints}).

The last two lines in (\ref{eq:H,H bracket of new constraints}) do not contain
$H_r$ as an overall factor, and thus they must vanish.  To simplify the
analysis we restrict ourselves to functions $B$ of the form
\begin{eqnarray}
    B = B (E^x , K_\varphi , (E^x)'/E^\varphi)
    \ .
    \label{eq:B coefficient - Linear combination}
\end{eqnarray}
All arguments of such a function are of density weight zero and therefore
fulfill the earlier condition (\ref{BHr}) on $B$.  With this choice, the
fourth line in (\ref{eq:H,H bracket of new constraints}) vanishes because of
antisymmetry of the Poisson bracket, and the last line can be written as
$H^{{\rm (old)}}[ F (N M' - M N')]$, where
\begin{equation}
  F = \frac{\partial}{\partial M'} \left( \not{\delta}_{B M , 0}^{{\rm (old)}}
    B \right) + A B = B\left(\frac{\partial}{\partial M'} \left( \not{\delta}_{M , 0}^{{\rm (old)}}
      B \right) + A\right)
\end{equation}
is independent of $M$ and $N$.
The condition $F = 0$ directly relates $A$ to $B$ via
\begin{eqnarray}
    A 
    &=& - \frac{\partial}{\partial M'} \left( \delta_{M , 0}^{{\rm (old)}} B \right)
    \ ,
    \label{eq:A coefficient - Linear combination}
\end{eqnarray}
which is of density weight $-1$, and therefore fulfills (\ref{eq:H,H_r bracket
  conditions}).

The first three lines in (\ref{eq:H,H bracket of new constraints}) then determine
the new structure function via $\{ H^{{\rm (new)}} [N] , H^{{\rm (new)}} [M]
\} = H_r [q^{x x}_{{\rm (new)}} (N M' - N' M)]$:
\begin{eqnarray}
    q^{x x}_{{\rm (new)}}
    &=& B^2 q^{x x}_{{\rm (old)}}
    + B \frac{\partial}{\partial M_1'} \left( \delta_{M_1 , 0}^{{\rm (old)}} A \right)
    \nonumber\\
    &&+ \left(
    \frac{\partial B}{\partial E^x} \frac{\partial A}{\partial K_x'}
    + \frac{\partial B}{\partial E^\varphi} \frac{\partial A}{\partial K_\varphi'}
    - \frac{\partial B}{\partial K_\varphi} \frac{\partial A}{\partial (E^\varphi)'}
    - \frac{\partial B}{\partial (E^x)'} \frac{\partial A}{\partial K_x}
    \right) H^{{\rm (old)}}
    \nonumber\\
    &&+ \frac{1}{2} \left(
    \frac{\partial A}{\partial E^x} \frac{\partial A}{\partial K_x'}
    + \frac{\partial A}{\partial E^\varphi} \frac{\partial A}{\partial K_\varphi'}
    - \frac{\partial A}{\partial K_x} \frac{\partial A}{\partial (E^x)'}
    - \frac{\partial A}{\partial K_\varphi} \frac{\partial A}{\partial (E^\varphi)'}
    \right) H_r
    \ ,
    \label{eq:Structure function - Linear combination}
\end{eqnarray}
where we have neglected possible second-order derivative terms in $A$, for
which a straightforward extension of the present analysis would be needed.
The covariance condition (\ref{eq:Covariance condition - modified -
  spherical}) applied to the structure function (\ref{eq:Structure function -
  Linear combination}) takes the form
\begin{eqnarray}
    \frac{\partial}{\partial (\epsilon^0)'} \delta_{\epsilon^0}^{{\rm (new)}}
  q^{x x}_{{\rm (new)}} \bigg|_{{\rm O.S.}}
    = \frac{\partial}{\partial (\epsilon^0)''} \delta_{\epsilon^0}^{{\rm
  (new)}} q^{x x}_{{\rm (new)}} \bigg|_{{\rm O.S.}}
    = \cdots
    = 0
    \ .
    \label{eq:Covariance condition - Linear combination}
\end{eqnarray}

\subsubsection{Constraints of the spherically symmetric theory}

Using the Hamiltonian constraints,
(\ref{eq:Hamiltonian constraint - spherical symmetry - Gauss solved}), and the
classical structure function, $q^{x x}_{{\rm (old)}} = E^x/(E^\varphi)^2$, the
anomaly-free linear combination of the constraints of the form 
(\ref{Hnew}) is obtained from (\ref{eq:B
  coefficient - Linear combination}) and (\ref{eq:A coefficient - Linear
  combination}):
\begin{eqnarray}
    A &=& - \frac{\sqrt{E^x} (E^x)'}{2 (E^\varphi)^2} \frac{\partial B}{\partial K_\varphi}
    - 2 K_\varphi \sqrt{E^x} \frac{\partial B}{\partial (E^x)'}
    \ .
    \label{eq:A coefficient - Linear combination - spherical - Anomaly-free}
\end{eqnarray}
The structure function (\ref{eq:Structure function - Linear combination}) of
the resulting anomaly-free brackets of hypersurface-deformation form then equals
\begin{eqnarray}
    q^{x x}_{{\rm (new)}} 
    &=& \frac{E^x}{(E^\varphi)^2} B^2
    - \frac{K_\varphi E^x}{(E^\varphi)^2} B \frac{\partial B}{\partial K_\varphi}
    - \left( \frac{\sqrt{E^x} (E^x)'}{2 (E^\varphi)^2} \right)^2 B \frac{\partial^2 B}{(\partial K_\varphi)^2}
    \nonumber\\
    &&+ \frac{E^x (E^x)'}{(E^\varphi)^2} B \frac{\partial B}{\partial (E^x)'}
    + \left( 2 K_\varphi \sqrt{E^x} \right)^2 B \frac{\partial^2 B}{(\partial (E^x)')^2}
    + \frac{2 K_\varphi E^x (E^x)'}{(E^\varphi)^2} B \frac{\partial^2 B}{\partial K_\varphi \partial (E^x)'}\,.
    \label{eq:Structure function - Linear combination - spherical - Anomaly-free}
\end{eqnarray}
(The second and third line of (\ref{eq:Structure function - Linear
  combination}) vanish identically in the present case.)

The covariance condition (\ref{eq:Covariance condition - Linear combination})
requires that
\begin{eqnarray}
    && 16 K_\varphi \left( 3 K_\varphi \partial_{z} B \left( 4 K_\varphi \partial_{z}^2 B + \partial_{K_\varphi} B \right)
    + B \left( K_\varphi \left( 4 K_\varphi \partial_{z}^3 B + 3 \partial_{K_\varphi} \partial_{z} B  \right) - 3 \partial_{z} B \right) \right)
    \nonumber\\
    &&+ 12 z \Bigg[
    K_\varphi \partial_{K_\varphi} B \left(4 K_\varphi \partial_{z}^2 B + \partial_{K_\varphi} B \right)
    + 4 K_\varphi \partial_{z} B \left( \partial_{z} B + 2 K_\varphi \partial_{K_\varphi} \partial_{z} B \right)
    \nonumber\\
    && \qquad
    + B \left( K_\varphi \left( 4 \partial_{z}^2 B + 4 K_\varphi \partial_{K_\varphi} \partial_z^2 B  + \partial_{K_\varphi}^2 B \right) - \partial_{K_\varphi} B \right)
    \Bigg]
    \nonumber\\
    &&+ 12 z^2 \Bigg[
    \partial_{K_\varphi} B \left( \partial_{z} B + 2 K_\varphi \partial_{K_\varphi} \partial_{z} B \right)
    + K_\varphi \partial_{z} B \partial_{K_\varphi}^2 B
    + B \left( \partial_{K_\varphi} \partial_{z} B + K_\varphi \partial_{K_\varphi}^2 \partial_z B \right)
    \Bigg]
    \nonumber\\
    &&+ z^3 \Bigg[
    3 \partial_{K_\varphi} B \partial_{K_\varphi}^2 B
    + B \partial_{K_\varphi}^3 B
    \Bigg]
    = 0 
    \ ,
    \label{eq:Covariance condition - linear combination - classical spherical}
\end{eqnarray}
where $z = (E^x)' / E^\varphi$.  If we further simplify the form of $B$ to
$B = B(K_\varphi,E^x)$, the long condition (\ref{eq:Covariance condition -
  linear combination - classical spherical}) reduces to two shorter
equations for $B$:
\begin{eqnarray}
    K_\varphi \left( \frac{\partial B}{\partial K_\varphi} \right)^2 + B \left(K_\varphi \frac{\partial^2 B}{(\partial K_\varphi)^2} - \frac{\partial B}{\partial K_\varphi} \right) &=& 0
    \\
    B \frac{\partial^3 B}{(\partial K_\varphi)^3} + 3 \frac{\partial B}{\partial K_\varphi} \frac{\partial^2 B}{(\partial K_\varphi)^2}
    &=& 0 
    \,.
\end{eqnarray}
These equations have the general solutions
$B = c_1 \sqrt{c_2 \pm K_\varphi^2}$ and
$B= \tilde{c}_1 \sqrt{ \tilde{c}_2 \pm K_\varphi^2 + \tilde{c}_3 K_\varphi}$,
respectively, where $c_i$ and $\tilde{c}_i$ are free functions of $E^x$.
Consistency between the two solutions yields
\begin{equation}
    B_s (K_\varphi , E^x) = \mu \sqrt{ 1 - s \lambda^2 K_\varphi^2}
                              \ ,\label{eq:Covariant linear combination - classical - vacuum}
\end{equation}
where $s = \pm 1$, $\mu = \mu (E^x) , \lambda = \lambda(E^x)$.
This result implies
\begin{equation}
    A_s = \mu \frac{\sqrt{E^x} (E^x)'}{2 (E^\varphi)^2} \frac{s \lambda^2
      K_\varphi}{\sqrt{1 - s \lambda^2 K_\varphi^2}}
\end{equation}
via (\ref{eq:A coefficient - Linear combination - spherical - Anomaly-free}), and
  \begin{equation}
    q^{x x}_{{\rm (new)}} = 
    \mu^2 \left( 1 + \frac{s \lambda^2}{1 - s \lambda^2 K_\varphi^2} \left( \frac{(E^x)'}{2 E^\varphi} \right)^2 \right) \frac{E^x}{(E^\varphi)^2}
\end{equation}
follows from (\ref{eq:Structure function - Linear combination - spherical -
  Anomaly-free}).  The classical constraint and structure function are
recovered in the limit $\lambda \to 0$, $\mu \to 1$. The new structure
function is always positive for $s=+1$ and therefore directly determines the
radial metric of an emergent line element.  (In this case, the modified theory
is equivalent to what has been analyzed in \cite{SphSymmEff,SphSymmEff2} for
constant $\lambda$ and a specific $\mu$ depending on $\lambda$. In these
papers, the covariance conditions had been checked specifically for the
modified constraints without a general underlying theory.)  For $s=-1$ there
may be regions of signature change.

The case of $s=1$ is interesting because the square root in (\ref{eq:Covariant
  linear combination - classical - vacuum}) then implies a bounded curvature
component $|K_\varphi| \leq 1 / \lambda$. The modified Hamiltonian constraint
in this case equals
\begin{eqnarray}
    H^{{\rm (new)}} &=& \mu \sqrt{1 - s \lambda^2 K_\varphi^2} \Bigg( \left( \frac{1}{8 \sqrt{|E^x|} E^\varphi} - s \lambda^2 \frac{\sqrt{E^x}}{2 (E^\varphi)^2} \frac{K_\varphi K_x}{1 - s \lambda^2 K_\varphi^2} \right) ((E^x)')^2
    \nonumber\\
    &&\qquad
    - \frac{\sqrt{|E^x|}}{2 (E^\varphi)^2} (E^x)' (E^\varphi)'
    + \frac{\sqrt{|E^x|}}{2 E^\varphi} (E^x)''
    + s \lambda^2 \frac{\sqrt{E^x}}{2 (E^\varphi)^2} \frac{E^\varphi K_\varphi}{1 - s \lambda^2 K_\varphi^2} (E^x)' K_\varphi'
    \nonumber\\
    &&\qquad
    - \frac{E^\varphi}{2 \sqrt{|E^x|}}
    - \frac{E^\varphi K_\varphi^2}{2 \sqrt{|E^x|}}
    - 2 K_\varphi \sqrt{|E^x|} K_x \Bigg)
    \ .
    \label{eq:Hamiltonian constraint - spherical symmetry - Covariant linear combination}
\end{eqnarray}
This constraint was first found in
\cite{SphSymmMatter,SphSymmMatter2}, up to a canonical transformation
\begin{eqnarray}
    K_\varphi &\to& \frac{\sin (\lambda K_\varphi)}{\lambda}
    \quad ,
    \quad
    E^\varphi \to \frac{E^\varphi}{\cos (\lambda K_\varphi)}
    \label{eq:Canonical transformation - holonomy model}
\end{eqnarray}
(for constant $\lambda$) that preserves the diffeomorphism constraint.

The canonically transformed constraint equals
\begin{eqnarray}
    H_{\rm c c} &=&
    - \mu \frac{\sqrt{E^x}}{2} \Bigg[
    \frac{E^\varphi}{E^x} \left( 1 + \frac{\sin^2 ( \lambda K_\varphi)}{\lambda^2} \right)
    + 4 K_x \frac{\sin (2 \lambda K_\varphi)}{2 \lambda}
    \nonumber\\
    &&\qquad
    - \frac{((E^x)')^2}{4 E^\varphi} \left( \frac{1}{E^x} \cos^2 (\lambda K_\varphi)
    - \frac{K_x}{E^\varphi} 2 \lambda \sin (2 \lambda K_\varphi) \right)
    \nonumber\\
    &&\qquad
    + \cos^2 (\lambda K_\varphi) \frac{(E^x)' (E^\varphi)'}{(E^\varphi)^2}
    - \cos^2 (\lambda K_\varphi) \frac{(E^x)''}{E^\varphi}
    \Bigg]
    \ .
    \label{eq:Hamiltonian constraint - spherical symmetry - Covariant linear combination - Canonical transformation}
\end{eqnarray}
The structure function (\ref{eq:Covariant linear combination - classical -
  vacuum}) must then also be transformed, yielding
\begin{eqnarray}
    q^{x x}_{\rm cc}
    &=&
    \mu^2 \cos^2 (\lambda K_\varphi) \left( 1
    + \left( \frac{\lambda (E^x)'}{2 E^\varphi} \right)^2 \right)
    \frac{E^x}{(E^\varphi)^2}
    \ .
    \label{eq:Structure function - spherical symmetry - Covariant linear combination - Canonical transformation}
\end{eqnarray}
By construction, this function implies a covariant emergent line element.
Static space-time solutions of the dynamical equations generated by the
constraint (\ref{eq:Hamiltonian constraint - spherical symmetry - Covariant
  linear combination - Canonical transformation}) and the classical
diffeomorphism constraint have been studied in \cite{SphSymmEff,SphSymmEff2}
for $\mu=1/\sqrt{1+\lambda^2}$, where covariance was also demonstrated
explicitly.  In this space-time, the surface of maximum curvature
$K_{\varphi}$ is a surface of reflection symmetry, as is readily seen from
(\ref{eq:Hamiltonian constraint - spherical symmetry - Covariant linear
  combination - Canonical transformation}) and (\ref{eq:Structure function -
  spherical symmetry - Covariant linear combination - Canonical
  transformation}).

Physical properties of this modified space-time are independent of the
canonical transformation because the evaluations of dynamics and covariance
are based completely on Poisson brackets.  This observation makes it clear
that the curvature bound and the avoidance of the classical singularity in
\cite{SphSymmEff} must be a consequence of the phase-space dependent linear
combination of hypersurface-deformation generators, which can lead to modified
space-time solutions because it changes the emergent normal direction.

These physical implications are independent of the use of the canonical
transformation or periodic functions originally intended to model holonomies
of loop quantum gravity; they are more general properties of emergent modified
gravity. (The canonical transformation is non-bijective, which has been argued
in a different model to allow new physical effects \cite{CovPol}, but this
possibility has been ruled out by \cite{NonCovPol}.) Nevertheless, the
application of the canonical transformation (\ref{eq:Hamiltonian constraint -
  spherical symmetry - Covariant linear combination - Canonical
  transformation}) has a technical advantage because, by its non-bijective
nature, the holonomy-like variables can be extended to both sides of the
reflection-symmetry surface. The constraint (\ref{eq:Hamiltonian constraint -
  spherical symmetry - Covariant linear combination}) diverges at the
maximum-curvature surface, $\lambda K_\varphi \to 1$, while the constraint
(\ref{eq:Hamiltonian constraint - spherical symmetry - Covariant linear
  combination - Canonical transformation}) remains finite,
$H^{{\rm (new)}}_{c c} \to - \mu E^\varphi / \sqrt{E^x}$ as
$\lambda K_\varphi \to \pi / 2$. This behavior is possible because the
non-bijective canonical transformation maps all finite values of $E^{\varphi}$
to infinite values at $\lambda K_{\varphi}=\pi/2$, suppressing terms that
originally diverge as the maximum-curvature surface is approached. The
non-bijective nature of the canonical transformation therefore does have an
implication, but only on the convenient parameterization of the surface and
not on the surrounding space-time regions where $E^{\varphi}$ is finite in
both descriptions.

Our solution for $s=-1$ has not been found before. It may be transformed
  canonically as in the $s=+1$ case, using hyperbolic instead of trigonometric
  functions. There is no curvature bound in this case, but the possibility of
  signature change might turn it into an interesting model system. We leave a
  detailed analysis to future work.

\subsection{Off-shell Partial Abelianization}
\label{sec:Partial}

As another application of our general equations, we can systematically
rederive the partial Abelianization of spherically symmetric constraints from
\cite{LoopSchwarz,LoopSchwarz2}.  To this end, we need to find a function $B$
in (\ref{eq:Structure function - Linear combination - spherical -
  Anomaly-free}) that eliminates the structure function in an anomaly-free
way: $q^{x x}_{{\rm (new)}} \equiv q^{x x}_{{\rm (A)}} = 0$. According to
(\ref{eq:Structure function - Linear combination - spherical - Anomaly-free}),
this is possible if
\begin{equation}
    B = K_\varphi B_x(E^x)
  \end{equation}
with some function $B_x$ that depends only on $E^x$,  such that
  \begin{equation}
    A = - \frac{\sqrt{E^x} (E^x)'}{(E^\varphi)^2} B_x\,.
\end{equation}
This solution of the Abelianization condition $q^{x x}_{{\rm (new)}} = 0$ is
unique up to a choice of $B_x = B_x (E^x)$.  The resulting Abelianized
Hamiltonian constraint equals
\begin{eqnarray}
    \frac{H^{{\rm (A)}}}{B_x}
    &=& \bigg(\frac{((E^x)')^2}{8 \sqrt{|E^x|} E^\varphi} \left( 1 + 8 E^x K_x \right)
    - \frac{E^\varphi}{2 \sqrt{|E^x|}} \left( 1 + K_\varphi^2 \right)
    - 2 K_\varphi \sqrt{|E^x|} K_x 
    \nonumber\\
    &&- \frac{\sqrt{|E^x|} (E^x)' (E^\varphi)'}{2 (E^\varphi)^2} 
    + \frac{\sqrt{|E^x|} (E^x)''}{2 E^\varphi}
    \bigg) K_\varphi
    - \frac{\sqrt{E^x} (E^x)'}{E^\varphi} K_\varphi'
    \ .
    \label{eq:Hamiltonian constraint - spherical symmetry - Abelianized}
\end{eqnarray}

Compared with the constructions in \cite{LoopSchwarz,LoopSchwarz2}, our
results provide a local and off-shell pathway to partial Abelianizations
without the need of an additional integration by parts in the Hamiltonian
constraint, integrating the lapse function. Our method therefore supports one
of the motivations of \cite{LoopSchwarz,LoopSchwarz2}, which is to simplify
common quantization procedures that are often untractable in the presence of
structure functions. The fully local construction given here provides further
simplifications in a quantization procedure that also aims to supply solutions
of the Abelianized theory with space-time interpretations. However, space-time
considerations require a transformation back to brackets of
hypersurface-deformation form: A vanishing structure function, as found in a
partially Abelianized theory, makes it impossible to interpret solutions of
the theory as emergent space-times because the theory does not provide an
unambiguous choice of $q^{xx}_{\rm eff}$. Formally, our covariance condition
is trivially satisfied in this case, but only because there is no emergent
line element to begin with.

Instead of using a full quantization right away, one may begin with an
analysis of modifications that are sometimes necessary in certain quantization
approaches, such as ``polymerization'' or the substitution of periodic
functions for extrinsic-curvature components in models of loop quantum
gravity. To have a chance of being covariant, such modifications must be
compatible with suitable corresponding modifications to the
hypersurface-deformation constraint (\ref{eq:Hamiltonian constraint -
  spherical symmetry - Gauss solved}), in such a way that the latter still
satisfies the covariance condition (\ref{eq:Covariance condition - modified -
  spherical}). If one is interested in a space-time picture with an
emergent
line element, this condition remains in place also for the Abelian constraint
(\ref{eq:Hamiltonian constraint - spherical symmetry - Abelianized}).

As an example, we may use the constraint (\ref{eq:Hamiltonian
  constraint - spherical symmetry - Covariant linear combination - Canonical
  transformation}) and its structure function (\ref{eq:Structure function -
  spherical symmetry - Covariant linear combination - Canonical
  transformation}), already modelling holonomy modifications, and then perform
the partial Abelianization as done above using (\ref{eq:B coefficient - Linear
  combination}), (\ref{eq:A coefficient - Linear combination}), and
(\ref{eq:Structure function - Linear combination}).
Considering $B = B(E^x,K_\varphi)$, the resulting Abelianized constraint is
\begin{eqnarray}
    \frac{H^{{\rm (A)}}_{\rm c c}}{B_x} &=&
    - \frac{\sqrt{E^x}}{2} \Bigg[
     \frac{E^\varphi}{E^x} \left( 1 + \frac{\sin^2 ( \lambda K_\varphi)}{\lambda^2} \right) \frac{\tan (\lambda K_\varphi)}{\lambda}
    + 4 K_x \frac{\sin^2 (\lambda K_\varphi)}{\lambda^2}
    \nonumber\\
    &&\qquad
    - \frac{((E^x)')^2}{E^\varphi} \left( \frac{1}{4 E^x} \frac{\sin (2 \lambda K_\varphi)}{2 \lambda}
    - \frac{K_x}{E^\varphi} \lambda^2 \frac{\sin^2 (\lambda K_\varphi)}{\lambda^2}
    + \frac{K_x}{E^\varphi} \right)
    \nonumber\\
    &&\qquad
    + \sec (\lambda K_\varphi) \frac{\left(E^x\right)'}{E^\varphi} \left(\frac{\sin (\lambda K_\varphi)}{\lambda}\right)'
    + \frac{\sin (2 \lambda K_\varphi)}{2 \lambda} \left( \frac{(E^x)' (E^\varphi)'}{(E^\varphi)^2}
    - \frac{(E^x)''}{E^\varphi} \right)
    \Bigg]
    \ ,
    \label{eq:Hamiltonian constraint - spherical symmetry - Covariant linear combination - Canonical transformation - Abelian}
\end{eqnarray}
which is again unique up to a choice of $B_x = B_x (E^x)$.  This constraint
re-introduces the divergence at $\lambda K_\varphi = \pi / 2$ owing to the
function $\tan (\lambda K_\varphi)$.  The appearance of
$\tan (\lambda K_\varphi)$ in the first line and $\sec (\lambda K_\varphi)$ in
the $K_\varphi'$ term, complicates a promotion of $H^{{\rm (A)}}_{\rm c c}$
to an operator and a corresponding discussion of covariance at the full
quantum level.

Another key difference between (\ref{eq:Hamiltonian constraint - spherical
  symmetry - Covariant linear combination - Canonical transformation -
  Abelian}) and the Abelianized constraint of
\cite{LoopSchwarz,LoopSchwarz2} is the lack of $K_x$ in the latter,
which facilitates loop quantization as no radial holonomies are needed.
However, imposing the covariance condition, radial holonomy modifications are
not allowed in the present spherically reduced model, presenting an ongoing
challenge to a complete loop quantization of black holes. To see this, we will
analyze the general case of modified Hamiltonian constraints in spherically
symmetric models, which may describe combinations of possible covariant
versions of holonomy modifications with phase-space dependent linear
combinations of hypersurface-deformation generators.

\section{General modified Hamiltonian constraints}
\label{sec:Modified spherically symmetric theory General}

In the preceding section we have demonstrated that phase-space dependent
linear combinations of the constraints can give rise to modified gravity
theories. We used several simplifying assumptions, such as in the limited
dependence of one of the linear coefficients, $B$, on the phase-space fields,
for this demonstration. In the present section, we continue to work with the
same models, given by the phase space and diffeomorphism constraint of
spherical symmetry, but aim to derive a more general form of modified
Hamiltonian constraints consistent with general covariance according to our
new condition.

The results can be understood as modified theories of gravity in which the
Hamiltonian constraint may be subject to a number of different modifications,
motivated for instance by canonical approaches to quantum gravity. For full
generality, one should then also allow for possible phase-space dependent
linear combinations with the diffeomorphism constraint since it is not certain
that a theory of quantum space-time would follow the classical separation into
tangential and normal deformations of spacelike hypersurfaces. Such linear
combinations also provide additional free functions compared with
modifications of the Hamiltonian constraint by itself. As we will demonstrate,
these free functions help to regain general covariance in an emergent
space-time description of modified constraints.

In this way, we consider general modifications to the spherically symmetric
theory with canonical variables $(K_\varphi , E^\varphi)$ and $(K_x , E^x)$,
without introducing any additional degrees of freedom as they would be implied
by higher time derivatives in the action. We therefore explore possibilities
of modified gravity that do not require new degrees of freedom, reducing the
danger of instabilities that might otherwise arise as in higher-curvature
effective actions; see for instance \cite{OstrogradskiProblem}.

Once we modify the Hamiltonian constraint in a specific way, the constraint
brackets (\ref{eq:Hypersurface deformation algebra - spherical})--(\ref{eq:H,H
  bracket}) determine the radial metric component via
$\tilde{q}_{x x} = 1 / |\tilde{q}^{x x}|$. As before, the angular component of
the metric cannot be determined by the constraint brackets.  For now, we
include a generic expression
$\tilde{q}_{\vartheta \vartheta} = \tilde{q}_{\vartheta \vartheta} (E^x)$ for
the modified angular component. As before, the covariance condition for the
angular component of the emergent metric implies (\ref{eq:Covariance condition
  on K_x - modified - spherical}) and (\ref{eq:Covariance condition - modified
  - spherical}), using
$\delta_{\epsilon^0} E^x = - \delta \tilde{H} [\epsilon^0]/\delta K_x$.

The emergent space-time metric is then given by
\begin{equation}
    {\rm d} s^2 = - {\rm sgn}(\tilde{q}^{xx}) N^2 {\rm d} t^2 + |\tilde{q}_{x
      x}| ( {\rm d} x + N^r {\rm 
      d} t )^2 + \tilde{q}_{\vartheta \vartheta} {\rm d} \Omega^2
    \,.
\end{equation}
Since the sign of $\tilde{q}_{\vartheta\vartheta}$ is not strictly determined
within a spherically symmetric theory, we assume that this metric component
remains positive. The emergent space-time line element then follows the second
option presented in Section~\ref{sec:SigChange}, applied to the
$1+1$-dimensional radial space-time. In this way, the 4-dimensional space-time
line element that includes the angular term does not have a distinguished
spatial direction based on signature.

\subsection{Modified constraint brackets}
\label{sec:Modified constraint algebra}

We follow the procedure employed in \cite{SphSymmMatter,SphSymmMatter2}, but
consider an expanded version of the Hamiltonian constraint to include
solutions not contained in these papers. We use the general ansatz
\begin{eqnarray}
    H &=& a_0
    + ((E^x)')^2 a_{x x}
    + ((E^\varphi)')^2 a_{\varphi \varphi}
    + (E^x)' (E^\varphi)' a_{x \varphi}
    + (E^x)'' a_2
    \nonumber\\
    &&+ (K_\varphi')^2 b_{\varphi \varphi}
    + (K_\varphi)'' b_2
    + (E^x)' K_\varphi' c_{x \varphi}
    + (E^\varphi)' K_\varphi' c_{\varphi \varphi}
    + (E^\varphi)'' c_2
    \label{eq:Hamiltonian constraint ansatz - Generalized vacuum - Extended}
\end{eqnarray}
for our Hamiltonian constraint, where $a_0$, $a_{i j}$, $a_2$,
$b_{\varphi \varphi}$, $b_2$, $c_2$, $c_{i j}$ are all functions of the phase
space variables, but not of their derivatives. (Here and from now on, we drop
the tilde on $H$ with the understanding that we are dealing with modified
constraints.)  We have included terms quadratic in first-order radial
derivatives and linear in second-order radial derivatives of all the phase
space variables, except of $K_x$ because this would break covariance as
demanded by (\ref{eq:Covariance condition on K_x - modified -
  spherical}). Terms linear in $K_{\varphi}'$, with coefficient $c_{x\varphi}$
or $c_{\varphi\varphi}$, may be viewed as derivative corrections, or as
contributions from the diffeomorphism constraint in a phase-space dependent
linear combination with the Hamiltonian constraint.

Starting from this ansatz we will obtain the conditions for it to satisfy the
hypersurface-deformation brackets (\ref{eq:Hypersurface deformation algebra -
  spherical})--(\ref{eq:H,H bracket}), possibly with a modified
structure function $\tilde{q}^{x x}$.

\subsubsection{$\{ H , H_r\}$ bracket}

The bracket $\{ H[N] , H_r [N^r] \}$ can be written as
\begin{equation} \label{HHr}
    \{ H[N] , H_r [N^r] \}
    = \int {\rm d} x \ N^r \bigg[ N \mathcal{F}_0 + N' \mathcal{F}_1 + N'' \mathcal{F}_2 \bigg]
\end{equation}
using integration by parts to avoid derivatives of $N^r$.  For this result to
match (\ref{eq:H,H_r bracket}), we set $\mathcal{F}_1=\mathcal{F}_2=0$ and
$\mathcal{F}_1 + H = 0$. Since all the free functions in the Hamiltonian
constraint (\ref{eq:Hamiltonian constraint ansatz - Generalized vacuum -
  Extended}) are, by definition, independent of derivatives of the phase space
variables, any terms in the equations implied by (\ref{HHr}) that
multiply different kinds of derivatives must vanish independently.

For a generic Hamiltonian constraint as used here, there is a rather large
number of such equations; see Appendix~\ref{appHam} for more details. They
imply that the constraint must have the form
\begin{eqnarray}
    H &=& - \sqrt{E^x} \frac{g}{2} \bigg[ E^\varphi A_0
    + \frac{((E^x)')^2}{E^\varphi} A_{x x}
    + \frac{(E^x)' (E^\varphi)'}{(E^\varphi)^2}
    - \frac{(E^x)''}{E^\varphi}
    \nonumber\\
    && \qquad
    + \frac{(K_\varphi')^2}{E^\varphi} B_{\varphi \varphi}
    + \frac{(E^x)' K_\varphi'}{E^\varphi} C_{x \varphi}
    + \left( \frac{(E^\varphi)' K_\varphi'}{(E^\varphi)^2}
    - \frac{(K_\varphi)''}{E^\varphi} \right) C_{\varphi \varphi} \bigg]
    \ ,
    \label{eq:Hamiltonian constraint ansatz with correct algebra with H_r -
       Generalized vacuum - Extended} 
\end{eqnarray}
where $A_0$, $A_{i j}$, $B_{\varphi}$, $C_{i j}$, and $g$ are free functions
of $E^x$, $K_\varphi$, and $K_x / E^\varphi$. (The function $g$ has been
factored out for convenience.)

\subsubsection{$\{ H, H \}$ bracket}

The bracket $\{ H[N] , H [M] \}$ can be written as
\begin{equation}
    \{ H[N] , H [M] \}
    = \int {\rm d} x \ \bigg[ (N M' - M N') \left( \mathcal{G}_0 -
      \mathcal{G}_1' + \mathcal{G}_2'' \right) - (N M''' - M N''')
    \mathcal{G}_2 \bigg] \,,
\end{equation}
where we used several integration by parts. Specific expressions for
$\mathcal{G}_0$, $\mathcal{G}_1$ and $\mathcal{G}_2$ can be obtained from an
explicit calculation of the Poisson bracket of two Hamiltonian constraints.
At this stage, they are quite long, but some of the terms have direct
implications that simplify the allowed dependence of coefficients on
phase-space degrees of freedom. We will indicate the simplifying implications
first and then proceed to more complicated terms.

For the bracket to match (\ref{eq:H,H bracket}), we must set $\mathcal{G}_2=0$
and $\mathcal{G}\equiv \mathcal{G}_0 - \mathcal{G}_1' = H_r \tilde{q}^{x x}$
for some function $\tilde{q}^{x x}$ of density weight $-2$.  The first one of
these equations, collecting the highest derivative terms, implies
\begin{equation}
    C_{\varphi \varphi} g^2 \frac{E^x}{4 (E^\varphi)^3} \left( (E^x)' +
      K_\varphi' C_{\varphi \varphi} \right) 
    = 0
\end{equation}
which, for a nontrivial Hamiltonian constraint, is solved only by
$C_{\varphi \varphi}=0$.

The second equation, $\mathcal{G} = \tilde{q}^{x x} H_r$ for some
$\tilde{q}^{xx}$, can again be separated into terms multiplying different
derivatives of the phase space variables.  The terms
\begin{equation}\label{G}
\mathcal{G} \supset - \frac{1}{4}E^x g (\partial g/\partial K_x) (E^x)''' +
G^x K_x' + G_\varphi (E^\varphi)'\,,
\end{equation}
with 
\begin{equation} \label{G2}
    G^x = - \frac{E^{\varphi}}{K_x}G_\varphi = \frac{g^2 E^x}{4 (E^{\varphi})^2}
  \frac{\partial^2 A_0}{\partial (K_x/E^{\varphi})^2} 
    \,,
\end{equation}
are the only ones that cannot contribute to the diffeomorphism constraint and
must therefore vanish separately. The first term in (\ref{G}) immediately
implies that $g$ does not depend on $K_x$, and therefore it does not depend on
$E^{\varphi}$ either because such a dependence could occur only in the
combination $K_x/E^{\varphi}$.

In the remaining terms of (\ref{G}), $G^x$ and $G_{\varphi}$ are phase-space
functions that do not depend on spatial derivatives. Since they appear in the
bracket of two Hamiltonian constraints via terms with only two spatial
derivatives, one in $NM'-MN'$ and one in $K_x'$ or $(E^{\varphi})'$, they can
result only from a Poisson bracket of the first term in (\ref{eq:Hamiltonian
  constraint ansatz with correct algebra with H_r - Generalized vacuum -
  Extended}), proportional to $E^{\varphi}A_0$, with some of the other
terms. The presence of $(E^x)''$ in any Hamiltonian constraint that has the
correct classical limit implies that $A_0$ can be at most linear in $K_x$, as
also seen directly from (\ref{G2}),
such that any spatial derivative of $\{A_0,E^x\}$ taken after integrating by
parts no longer produces terms with $K_x'$.  To summarize this step, $g$ does
not depend on $K_x/E^{\varphi}$ and $A_0$ is linear in $K_x/E^{\varphi}$:
\begin{equation}
    g = g (E^x , K_\varphi) \quad\mbox{and}\quad
    A_0 = f_0 + \frac{K_x}{E^\varphi} f_1
\end{equation}
where $f_0$ and $f_1$ are free functions of $E^x$ and $K_\varphi$.

The remaining non-zero terms in $\mathcal{G}$ . They are of the form
\begin{eqnarray}
    \mathcal{G} &=&
    G^\varphi K_\varphi' + G_x (E^x)'\label{Gexp} \\
    &&+ \left( F{^\varphi_x} (K_\varphi)' + F_{x x} (E^x)' \right) (E^x)''
    \nonumber\\
    &&+ \left(
    F^{\varphi \varphi} K_\varphi'
    +  F{_x^{2\varphi}} (E^x)'
    \right) K_\varphi''
    \nonumber\\
    &&+ \left( G^{\varphi \varphi x} (K_\varphi')^2
    + G^{x \varphi}_x K_\varphi' (E^x)'
    + G_{x x}^x ((E^x)')^2 \right) K_x'
    \nonumber\\
    &&
    + \left( G^{\varphi \varphi}_\varphi (K_\varphi')^2
    + G_{x \varphi}^\varphi (E^x)' K_\varphi'
    + G_{x x \varphi} ((E^x)')^2 \right) (E^\varphi)'
    \nonumber\\
    &&+ G^{\varphi \varphi \varphi} (K_\varphi')^3
    + G^{\varphi \varphi}_x (K_\varphi')^2 (E^x)'
    + G_{x x}^\varphi ((E^x)')^2 K_\varphi'
    + G_{x x x} ((E^x)')^3\,.
       \nonumber
\end{eqnarray}
Explicit expressions for all coefficients are given in Appendix~\ref{app:HH}.
All terms can in principle contribute to the diffeomorphism constraint. It is
therefore convenient to rearrange the terms according to
\begin{eqnarray}
  {\cal G}  &=&
    \bigg( \tilde{q}_0 + \tilde{q}_{2 x} (E^x)'' + \tilde{q}_{2 \varphi}
         K_\varphi'' + \left( {\tilde{q}_{1}}^{\varphi x} K_\varphi' +
         {\tilde{q}_1}\mbox{}_{x}^{x} (E^x)' \right) K_x' + \left(
         {\tilde{q}_1}\mbox{}^\varphi_\varphi K_\varphi' + {\tilde{q}_1}\mbox{}_{x \varphi}
         (E^x)' \right) (E^\varphi)' 
    \nonumber\\
    &&\qquad
    + {\tilde{q}_2}\mbox{}_{\varphi \varphi} (K_\varphi')^2
    + {\tilde{q}_2}\mbox{}_{x \varphi} (E^x)' K_\varphi'
    + {\tilde{q}_2}\mbox{}_{x x} ((E^x)')^2
    \bigg) H_r
    \,,
    \label{eq:G function - Generalized vacuum - Extended}
\end{eqnarray}
where all the $q$-coefficients contribute to the structure function of the
resulting hypersurface-deformation bracket.

In order to obtain the reqiored factorization as a multiple of the
diffeomorphism constraint, the terms in $\mathcal{G}$ must satisfy the
relations
\begin{equation}
    \frac{G^\varphi}{K_\varphi} = - \frac{G_x}{K_x}
    \ ,
    \label{eq:H,H bracket equations - a}
  \end{equation}
  \begin{equation}
    \frac{F{^\varphi_x}}{E^\varphi} = - \frac{F_{x x}}{K_x}
    \ ,
    \label{eq:H,H bracket equations - b}
  \end{equation}
  \begin{equation}
    \frac{F^{\varphi \varphi}}{E^\varphi} = - \frac{F{_x^{2\varphi}}}{K_x}
    \ ,
    \label{eq:H,H bracket equations - c}
  \end{equation}
  \begin{equation}
    \frac{G^{\varphi \varphi x}}{E^\varphi} = - a \frac{G^{x \varphi}_x}{K_x}
    \ ,
    \label{eq:H,H bracket equations - d}
  \end{equation}
  \begin{equation}
    (1-a) \frac{G^{x \varphi}_x}{E^\varphi} = - \frac{G^{x}_{x x}}{K_x}
    \ ,
    \label{eq:H,H bracket equations - e}
  \end{equation}
  \begin{equation}
  \frac{G^{\varphi \varphi}_\varphi}{E^\varphi} = - b \frac{G^{\varphi}_{x \varphi}}{K_x}
    \ ,
    \label{eq:H,H bracket equations - f}
  \end{equation}
  \begin{equation}
    (1-b) \frac{G^{\varphi}_{x \varphi}}{E^\varphi}= - \frac{G_{x x \varphi}}{K_x}
    \ ,
    \label{eq:H,H bracket equations - g}
  \end{equation}
  \begin{equation}
    \frac{G^{\varphi \varphi \varphi}}{E^\varphi} = - a_1 \frac{G^{\varphi \varphi}_x}{K_x}
    \ ,
    \label{eq:H,H bracket equations - h}
  \end{equation}
  \begin{equation}
    \tilde{b}_1 \frac{G_{x x}^\varphi}{E^\varphi}
    = - \frac{G_{x x x}}{K_x}
    \ ,
    \label{eq:H,H bracket equations - i}
  \end{equation}
  \begin{equation}
    (1-a_1) \frac{G^{\varphi \varphi}_x}{E^\varphi} = - (1-\tilde{b}_1) \frac{G_{x x}^\varphi}{K_x}
    \ ,
    \label{eq:H,H bracket equations - j}
\end{equation}
where $a$, $a_1$ and $\tilde{b}_1$ are arbitrary functions of $E^x$,
$K_\varphi$, and $K_x / E^\varphi$.  The first three equations can be solved
for some of the coefficients in (\ref{eq:Hamiltonian constraint ansatz with
  correct algebra with H_r - Generalized vacuum - Extended}):
\begin{eqnarray}
    B_{\varphi \varphi} 
    &=& \frac{1}{z} \frac{1}{2 f_0} \left( \frac{\partial f_0}{\partial K_\varphi} 
    + 2 f_0 \frac{\partial \ln g}{\partial K_\varphi}
    - \frac{\partial f_1}{\partial E^x}
    + 2 f_1 f_2
        \right)\nonumber\\
  &&
    + \frac{1}{z^2} \left( A_{x x} - f_2 \right)
    \ ,
    \label{eq:Anomlay-freedom equation simplified - 0 order - Generalized vacuum - Extended}\\
    C_{x \varphi}
    &=& - \frac{\partial \ln g}{\partial K_\varphi}
    - \frac{2}{z} ( A_{x x} - f_2 )
    \label{eq:Anomlay-freedom equation simplified - First 1&2 order - Generalized vacuum - Extended}
\end{eqnarray}
where $z = K_x/E^\varphi$ and $f_2 = f_2 (E^x , K_\varphi)$.
The remaining relations remain quite long and complicated.

The structure function is now given by
\begin{equation}
    \tilde{q}^{x x} =
    q_0
    + q_{2 x}
    + q_{2 \varphi}
    + {q_{1}}^x
    + {q_1}_\varphi
    + q_3
    \label{eq:Structure function of anomaly-free algrebra - Generalized vacuum - Extended}
\end{equation}
with
\begin{eqnarray}
    q_0 &=& \frac{G^\varphi}{E^\varphi}
    = \frac{E^x g^2}{4 (E^\varphi)^2}
    \left(
    2 f_0 B_{\varphi \varphi}
    - f_1 C_{x \varphi}
    + \frac{\partial f_1}{\partial K_\varphi}
    \right)
    \ ,
    \label{eq:Structure function - 0 order - Generalized vacuum - Extended}\\
    q_{2 x} &=& \frac{F{^\varphi_x}}{E^\varphi} (E^x)''
    = \frac{E^x g^2 (E^x)''}{4 (E^\varphi)^4 z^2} \frac{\partial C_{x \varphi}}{\partial z}
    \ ,
    \label{eq:Structure function simplified - First 1&2 order - Generalized vacuum - Extended}\\
    q_{2 \varphi} &=& \frac{F^{\varphi \varphi}}{E^\varphi} K_\varphi''
    = \frac{E^x g^2 K_\varphi''}{2 (E^\varphi)^4} \frac{\partial B_{\varphi \varphi}}{\partial z}
    \ .
    \label{eq:Structure function - Second 1&2 order - Generalized vacuum - Extended}\\
    {q_1}^x &=& \left( \frac{G^{\varphi \varphi x}}{E^\varphi} K_\varphi' - \frac{G^{x}_{x x}}{K_x} (E^x)' \right) K_x'
    \nonumber\\
    &=&
    \frac{E^x g^2 K_x'}{2 (E^\varphi)^6 z^4}
    \left(
    A_{x x}
    - f_2
    - z \frac{\partial A_{x x}}{\partial z}
    + \frac{z^2}{2} \frac{\partial^2 A_{x x}}{\partial z^2}
    \right)
    \left( \frac{a}{1- a} E^\varphi K_\varphi' - K_x (E^x)' \right)
    \ .
    \label{eq:Structure function - K_x' term - Generalized vacuum - Extended}\\
    {q_1}_\varphi
    &=& \left( \frac{G^{\varphi \varphi}_\varphi}{E^\varphi} K_\varphi'
    - \frac{G_{x x \varphi}}{K_x} (E^x)' \right) (E^\varphi)'
    = \frac{E^x g^2 (E^\varphi)'}{4 (E^\varphi)^6 z} \frac{\partial^2 A_{x x}}{\partial z^2} \left( \frac{b}{1 - b} E^\varphi K_\varphi'
    - K_x (E^x)' \right)
    \label{eq:Structure function - (E^varphi)' term - Generalized vacuum -
        Extended}
\end{eqnarray}
and
\begin{eqnarray}
    {q_3} &=& 
    \frac{G^{\varphi \varphi \varphi}}{E^\varphi} (K_\varphi')^2
    - b_1 \frac{G^{\varphi}_{x x}}{K_x} (E^x)' K_\varphi'
    - \frac{G_{x x x}}{K_x} ((E^x)')^2
    \nonumber\\
    &=& - \frac{E^x g^2}{2 (E^\varphi)^6 z^4} \bigg(
    A_{x x} ( A_{x x} - f_2 )
    - \frac{\partial}{\partial E^x} ( A_{x x} - f_2 )
    + \frac{z^2}{2} \frac{\partial \ln g}{\partial K_\varphi} \frac{\partial A_{x x}}{\partial z}
    \nonumber\\
    &&+ \frac{z}{2} \left( 
    \frac{\partial \ln g}{\partial K_\varphi} \frac{\partial \ln g}{\partial E^x}
    - \frac{1}{g} \frac{\partial^2 g}{\partial K_\varphi \partial E^x}
    + \frac{\partial^2 A_{x x}}{\partial E^x \partial z}
    - 2 f_2 \frac{\partial A_{x x}}{\partial z}
    - \frac{\partial A_{x x}}{\partial K_\varphi}
    \right)
    \bigg)
    \nonumber\\
    &&\times
    \left(
    \frac{a_1}{1 - a_1} \frac{1 - \tilde{b}_1}{\tilde{b}_1} (E^\varphi)^2 (K_\varphi')^2
    - \frac{1 - \tilde{b}_1}{\tilde{b}_1} E^\varphi K_x K_\varphi' (E^x)'
    + K_x^2 ((E^x)')^2 \right)
    \,.
    \label{eq:Structure function - K_varphi'(E^x)' term - Generalized vacuum - Extended}
\end{eqnarray}

\subsection{Covariance condition}

Now that we have the structure function (\ref{eq:Structure function of
  anomaly-free algrebra - Generalized vacuum - Extended}), we can apply the
covariance condition (\ref{eq:Covariance condition - modified - spherical})
which in our case is non-trivial up to the third-order derivative of the gauge
function.

The covariance condition must be evaluated on-shell.  In particular, one has
to pay attention to the on-shell property $H_r = 0$, which implies
$E^\varphi K_\varphi' = K_x (E^x)'$ and can mix some derivative terms that
were independent in the off-shell treatment so far.  In order to solve all the
covariance conditions, it is best to focus first on the highest derivative
terms of each one.

The highest-order derivative term of the first covariance condition is
\begin{eqnarray}
    \frac{\partial (\delta_\epsilon q^{x x})}{\partial (\epsilon^0)'}
  \bigg|_{\rm O.S.}
    \supset
    \frac{g^3 (E^x)^{3/2}}{8 (E^\varphi)^3 K_x^3 f_0} (E^x)''' \Bigg(
    2 g f_2 \left( 2 f_0 - z f_1 \right)
    - 4 f_0 g A_{x x}
    - 2 f_0 z \frac{\partial g}{\partial K_\varphi}
    \nonumber\\
    + g z \left( 2 f_0 \frac{\partial A_{x x}}{\partial z}
    + \frac{\partial f_1}{\partial E^x}
    - \frac{\partial f_0}{\partial K_\varphi} \right)
    \Bigg)
    \,.
\end{eqnarray}
The on-shell condition does not affect this equation, which has the solution
\begin{equation}
    A_{x x} = f_2
    - \frac{z}{2 f_0} \left( 2 f_0 \frac{\partial \ln g}{\partial K_\varphi}
    + \frac{\partial f_0}{\partial K_\varphi} 
    - \frac{\partial f_1}{\partial E^x}
    + 2 f_1 f_2 \right)
    + z^2 f_3
    \label{eq:Covariance condition - highest derivative term - (epsilon^0)' - Generalized vacuum - Extended}
\end{equation}
where $f_3 = f_3 (E^x,K_\varphi)$.
With this result for $A_{x x}$, the third covariance condition,
\begin{eqnarray}
    \frac{\partial (\delta_\epsilon q^{x x})}{\partial (\epsilon^0)'''}
    &=&
    \frac{g^3 (E^x)^{3/2}}{4 (E^\varphi)^2 K_x^4} \Bigg(
    - (E^x)' \left(
    2 z A_{x x}
    - 2 z f_2
    + z^2 \frac{\partial \ln g}{\partial K_\varphi}
    + \frac{z^2}{2 f_0} \left( 2 f_1 f_2 + \frac{\partial f_0}{\partial K_\varphi} - \frac{\partial f_1}{\partial E^x} \right)
    - z^2 \frac{\partial A_{x x}}{\partial z}
    \right)
    \nonumber\\
    &&+ \left( \frac{a}{1-a} K_\varphi' - z (E^x)' \right)
    \left(
    A_{x x}
    - f_2
    - z \frac{\partial A_{x x}}{\partial z}
    + \frac{z^2}{2} \frac{\partial^2 A_{x x}}{(\partial z)^2}
    \right)
    \Bigg)
    \, ,
\end{eqnarray}
vanishes.
The highest-order derivative term of the second covariance condition is
\begin{equation}
    \frac{\partial (\delta_\epsilon q^{x x})}{\partial (\epsilon^0)''}
    \supset
    \frac{g (E^x)^{3/2}}{E^\varphi} \left( E^\varphi K_\varphi'' - K_x (E^x)'' \right) f_3
    \,,
\end{equation}
and we conclude that $f_3 = 0$.

We now go back to some of the anomaly-freedom equations.  We note that we can
write equation (\ref{eq:H,H bracket equations - i}) in powers of $z$, which
terminates at second order because of the quadratic dependence on $K_x$. The
non-vanishing powers are
\begin{equation}
    z G^\varphi_{x x} =
    z G^{\varphi \ (1)}_{x x}
\end{equation}
and
\begin{equation}
    G_{x x x} =
    G^{(0)}_{x x x}
    + z G^{(1)}_{x x x}
\end{equation}
where all $G^{\varphi \ (i)}_{x x}$ and $G^{(i)}_{x x x}$ are independent of
$z$.  Thus, the anomaly-freedom equation (\ref{eq:H,H bracket equations - i})
requires $\tilde{b_1}=1$ and $G^{(0)}_{x x x} = 0$, which in turn implies
\begin{equation}
    0
    =
    f_0^2 \left[ \frac{\partial}{\partial E^x} \left( \frac{\partial \ln g}{\partial K_\varphi} - C_{x \varphi} \right)
    + 2 \frac{\partial f_2}{\partial K_\varphi} \right]
    \,.
    \label{eq:Anomaly-freedom equation (c) updated by covariance (epsilon^0)''' term - K_varphi'(E^x)' term - Generalized vacuum - Extended}
\end{equation}
We note that with $\tilde{b}_1=1$, the function $a_1$ drops out of all
equations.  Using these results we obtain $B_{\varphi \varphi}=0$, and the
anomaly-freedom equations~(\ref{eq:H,H bracket equations - d}), (\ref{eq:H,H
  bracket equations - e}), (\ref{eq:H,H bracket equations - f}), (\ref{eq:H,H
  bracket equations - g}), (\ref{eq:H,H bracket equations - h}), and
(\ref{eq:H,H bracket equations - j}) are automatically satisfied.

The remaining equations are given by lower-order derivatives of $E^x$ in

\[
  \frac{\partial}{\partial (\epsilon^0)'} \delta_{\epsilon} q^{x x} |_{\rm O.S.} =
  0\,,
\]
which has only two non-zero terms, with $(E^x)'$ and with $((E^x)')^3$, that
must vanish separately.  The vanishing of the $(E^x)'$-term implies
\begin{eqnarray}
    0 &=&
    \frac{\partial^2 \ln g}{(\partial K_\varphi)^2} f_1
    - 2 \frac{\partial \ln g}{\partial K_\varphi} \frac{\partial f_1}{\partial K_\varphi}
    - \frac{\partial^2 f_1}{(\partial K_\varphi)^2}\label{eq:Covariance equation - (epsilon^0)' term - (E^x)' term - Generalized vacuum - Extended}
    \\
    &&- \frac{\partial ( f_1 C_{x \varphi})}{\partial K_\varphi}
    + 3 f_1 \left( C_{x \varphi}^2
    + \frac{\partial C_{x \varphi}}{\partial K_\varphi}
    + \frac{\partial \ln g}{\partial K_\varphi} C_{x \varphi} \right)
    \,,
    \nonumber
\end{eqnarray}
and the vanishing of the $((E^x)')^3$-term implies
\begin{eqnarray}
    0
    &=&
    \left( \frac{\partial \ln g}{\partial K_\varphi}
    + 3 C_{x \varphi} \right) \left( \frac{\partial^2 \ln g}{(\partial K_\varphi)^2}
    + C_{x \varphi}^2
    + \frac{\partial C_{x \varphi}}{\partial K_\varphi}
    + \frac{\partial \ln g}{\partial K_\varphi} C_{x \varphi} \right)
    \nonumber\\
    &&+ \frac{\partial}{\partial K_\varphi} \left( \frac{\partial^2 \ln g}{\partial K_\varphi^2}
    + C_{x \varphi}^2
    + \frac{\partial C_{x \varphi}}{\partial K_\varphi}
    + \frac{\partial \ln g}{\partial K_\varphi} C_{x \varphi} \right)
    \,.
    \label{eq:Covariance equation - (epsilon^0)' term - ((E^x)')^3 term - Generalized vacuum - Extended}
\end{eqnarray}

This exhausts all the anomaly-freedom equations (\ref{eq:H,H bracket equations
  - a})--(\ref{eq:H,H bracket equations - j}).
Using these values and condition (\ref{eq:Anomaly-freedom equation (c) updated
  by covariance (epsilon^0)''' term - K_varphi'(E^x)' term - Generalized
  vacuum - Extended}), one can check that the covariance condition
$(\partial\delta_{\epsilon} q^{x x}/\partial (\epsilon^0)'')  |_{\rm O.S.} =
0$ is automatically satisfied.

To summarize, the general form of an anomaly-free and covariant Hamiltonian constraint is
\begin{eqnarray}
    H
    &=& - \sqrt{E^x} \frac{g}{2} \bigg[ E^\varphi \left( f_0 + \frac{K_x}{E^\varphi} f_1 \right)
    + \frac{((E^x)')^2}{E^\varphi} \left( f_2
    -  \frac{1}{2}\frac{K_x}{E^\varphi} \left( \frac{\partial \ln g}{\partial K_\varphi}
    + C_{x \varphi} \right) \right)
    \nonumber\\
    &&\qquad
    + \frac{(E^x)' (E^\varphi)'}{(E^\varphi)^2}
    - \frac{(E^x)''}{E^\varphi}
    + \frac{(E^x)' K_\varphi'}{E^\varphi} C_{x \varphi}
    \bigg]
    \label{eq:Hamiltonian constraint - Anomaly-free - covariant - spherical - vacuum - general - extended}
\end{eqnarray}
where $C_{x \varphi}$ is given by
\begin{equation}
    C_{x \varphi}
    = \frac{1}{f_0} \left( f_0 \frac{\partial \ln g}{\partial K_\varphi}
    + \frac{\partial f_0}{\partial K_\varphi} 
    - \frac{\partial f_1}{\partial E^x}
    + 2 f_1 f_2 \right)
    \ ,
    \label{eq:Anomlay-free, covariant extra coefficient - Generalized vacuum - Extended}
\end{equation}
and $g$, $f_0$, $f_1$, and $f_2$ are functions of $E^x$ and $K_\varphi$ that
must satisfy the equations (\ref{eq:Anomaly-freedom equation (c) updated by
  covariance (epsilon^0)''' term - K_varphi'(E^x)' term - Generalized vacuum -
  Extended}), (\ref{eq:Covariance equation - (epsilon^0)' term - (E^x)' term -
  Generalized vacuum - Extended}) and (\ref{eq:Covariance equation -
  (epsilon^0)' term - ((E^x)')^3 term - Generalized vacuum - Extended}).  The
structure function is
\begin{equation}
    \tilde{q}^{x x} =
    \left(
    \frac{\partial f_1}{\partial K_\varphi}
    - f_1 C_{x \varphi}
    - \frac{1}{2}
    \left(
    \frac{\partial^2 \ln g}{(\partial K_\varphi)^2}
    + C_{x \varphi}^2
    + \frac{\partial C_{x \varphi}}{\partial K_\varphi}
    + \frac{\partial \ln g}{\partial K_\varphi} C_{x \varphi}
    \right)
    \left( \frac{(E^x)'}{E^\varphi} \right)^2
    \right)
    \frac{g^2}{4}
    \frac{E^x}{(E^\varphi)^2}
    \,.
    \label{eq:Structure function - Anomaly-free - covariant - spherical - vacuum - general - extended}
\end{equation}
Unlike the structure function found in the preceding section by using only a
phase-space dependent linear combination of the constraints but no further
modification, this structure function is not guaranteed to be
positive. Suitable sign choices are therefore necessary when using the inverse
of this structure function in an emergent space-time line element, as
discussed in Section~\ref{sec:SigChange}.

\subsection{Applying canonical transformations}

By directly solving the required conditions, it can be shown that the set of
canonical transformations preserving the diffeomorphism constraint
(\ref{eq:Diffeomorphism constraint - spherical symmetry- Gauss solved}) and
leaving the variable $E^x$ invariant must have the form
\begin{eqnarray}
    K_\varphi &=& f_c (\tilde{E}^x , \tilde{K}_\varphi)
    \quad,\quad
    E^\varphi = \tilde{E}^\varphi \left( \frac{\partial f_c}{\partial \tilde{K}_\varphi} \right)^{-1}
    \ ,
    \nonumber\\
    K_x &=& \tilde{K}_x + \tilde{E}^\varphi \frac{\partial f_c}{\partial \tilde{E}^x} \left( \frac{\partial f_c}{\partial \tilde{K}_\varphi} \right)^{-1}
    \quad,\quad
    E^x = \tilde{E}^x
    \label{eq:Diffeomorphism-constraint-preserving canonical transformations - Spherical}
\end{eqnarray}
where the new variables are written with a tilde.  This canonical
transformation can be generalized by noting that the transformation
\begin{equation}
K_x = \frac{\partial (\alpha^2 E^x)}{\partial E^x} \tilde{K}_x\quad,\quad
\tilde{E}^x = \alpha^2 E^x
\end{equation}
is canonical and preseves the diffeomorphism constraint too, where
$\alpha = \alpha (E^x)$.
We will be using a combination of these as a subset of all
diffeomorphism-preserving canonical transformations:
\begin{eqnarray}
    K_\varphi &=& f_c (E^x , \tilde{K}_\varphi)\nonumber\\
    E^\varphi &=& \tilde{E}^\varphi \left( \frac{\partial f_c}{\partial
                  \tilde{K}_\varphi} \right)^{-1} 
    \nonumber\\
    K_x &=& \frac{\partial (\alpha_c^2 E^x)}{\partial E^x} \tilde{K}_x +
            \tilde{E}^\varphi \frac{\partial f_c}{\partial E^x} \left(
            \frac{\partial f_c}{\partial \tilde{K}_\varphi}
            \right)^{-1}\nonumber\\ 
    \tilde{E}^x &=& \alpha_c^2 (E^x) E^x
    \label{eq:Diffeomorphism-constraint-preserving canonical transformations - Spherical - general}
\end{eqnarray}
with type-3 generating function
\begin{equation}
    F_3 (E^x , E^\varphi , \tilde{K}_\varphi , \tilde{K}_x)
    =
    - f_c (E^x , \tilde{K}_\varphi) E^\varphi
    - \alpha_c^2 (E^x) E^x \tilde{K}_x
    \,.
\end{equation}

Let us now consider the Hamiltonian constraint (\ref{eq:Hamiltonian constraint
  - Anomaly-free - covariant - spherical - vacuum - general - extended}) and
perform a canonical transformation of the form
(\ref{eq:Diffeomorphism-constraint-preserving canonical transformations -
  Spherical - general}).  By focusing on the $(E^x)' (E^\varphi)'$ and
$(E^x)''$ terms, we see that the global factor transforms from
$g(E^x,K_\varphi)$ to $g_{\rm c c} (E^x,K_\varphi)$, where
\begin{equation}
    g_{\rm cc} (E^x , K_\varphi) =  g_{{\rm (old)}} (E^x / \alpha_c^2 , f_c ) \left( 1 - \frac{\partial \ln \alpha_c^2}{\partial E^x} \right) \frac{1}{\alpha_c^{3}} \frac{\partial f_c}{\partial K_\varphi}
    \,.
    \label{eq:Canonical transformation of global factor - Generalized vacuum - Extended}
\end{equation}
We then see that the $C_{x \varphi} (E^x , K_\varphi)$ coefficient transforms to
\begin{eqnarray}
   && C_{x \varphi}^{\rm cc} (E^x , K_\varphi)\label{eq:Canonical transformation of C_xphi coefficient - Generalized vacuum - Extended}\\
    &=& - \left( \frac{\partial f_c}{\partial K_\varphi} \right)^{-1}
    \frac{\partial^2 f_c}{\partial K_\varphi^2} 
    + \frac{\partial f_c}{\partial K_\varphi} C_{x \varphi}^{{\rm (old)}}
    (E^x / \alpha_c^2 , f_c (E^x , K_\varphi)) 
    \,. \nonumber
    \end{eqnarray}
By setting $f_c = f_c (K_\varphi)$ and $\alpha_c = 1$, it is therefore
possible to find, at least locally in phase space, a diffeomorphism-preserving
canonical transformation by solving an ordinary differential equation, such
that $C_{x \varphi}^{\rm cc}=0$.  After such a canonical transformation, the
Hamiltonian constraint and the structure function simplify by setting
$C_{x \varphi}= 0$.

Moreover, any modified angular component of the form
$\tilde{q}_{\vartheta \vartheta} = \alpha_c^{-2} (E^x) E^x$ can be mapped to
its classical form, $\tilde{q}_{\vartheta \vartheta} \to E^x$, by using the
canonical transformation (\ref{eq:Diffeomorphism-constraint-preserving
  canonical transformations - Spherical - general}) with $f_c=K_{\varphi}$ and the
necessary $\alpha_c$, which preserves $C_{x \varphi}=0$. 
With these choices, the residual canonical transformation has the form
\begin{eqnarray}
    K_\varphi &\to& f_x  K_\varphi
    - \tilde{\mu}_\varphi
    \quad,\quad
    E^\varphi \to \frac{E^\varphi}{f_x}
    \,,    \label{eq:Diffeomorphism-constraint-preserving canonical transformations - Spherical - residual after modulo}\\
    K_x &\to& K_x + E^\varphi K_\varphi \left( \frac{\partial \ln f_x}{\partial E^x}
    - \frac{1}{f_x} \frac{\partial \tilde{\mu}_\varphi}{\partial E^x} \right)
    \quad,\quad
    E^x \to E^x
    \,, \nonumber
\end{eqnarray}
where $f_x = f_x (E^x)$ and $\tilde{\mu}_\varphi = \tilde{\mu}_\varphi (E^x)$.

Because the anomaly-freedom equations and the covariance condition are all
based on Poisson brackets, canonical transformations leave them
form-invariant.  Therefore, $C_{x \varphi} = 0$ becomes a new condition on the
free functions through (\ref{eq:Anomlay-free, covariant extra coefficient -
  Generalized vacuum - Extended}), greatly simplifying the remaining
equations.
This observation allows us to obtain exact solutions to all anomaly-freedom
and covariance conditions as follows: We first solve
equation~(\ref{eq:Covariance equation - (epsilon^0)' term - ((E^x)')^3 term -
  Generalized vacuum - Extended}) for $g$, then solve
equation~(\ref{eq:Covariance equation - (epsilon^0)' term - (E^x)' term -
  Generalized vacuum - Extended}) for $f_1$, equation~(\ref{eq:Anomaly-freedom
  equation (c) updated by covariance (epsilon^0)''' term - K_varphi'(E^x)'
  term - Generalized vacuum - Extended}) for $f_2$, and
equation~(\ref{eq:Anomlay-free, covariant extra coefficient - Generalized
  vacuum - Extended}), set equal to zero, for $f_0$.

For future convenience we write the residual canonical transformation of all
terms in the Hamiltonian constraint according to 
(\ref{eq:Diffeomorphism-constraint-preserving canonical transformations -
  Spherical - residual after modulo}) as
\begin{eqnarray}\label{eq:Diffeomorphism-constraint-preserving canonical transformations of Constraint terms - Spherical - residual after modulo}
    g_{\rm cc} &= & g (E^x , f_x K_\varphi - \tilde{\mu}_\varphi) f_x
    \,,\\
    g_{\rm c c} f_0^{\rm cc}
    &=& g (E^x , f_x K_\varphi - \tilde{\mu}_\varphi) \left( \frac{f_0 (E^x , f_x K_\varphi - \tilde{\mu}_\varphi)}{f_x}
    + \left( \frac{\partial \ln f_x}{\partial E^x}
    - \frac{1}{f_x} \frac{\partial \tilde{\mu}_\varphi}{\partial E^x} \right) K_\varphi f_1 (E^x , f_x K_\varphi - \tilde{\mu}_\varphi) \right)
    \,,\\
    g_{\rm cc} f_1^{\rm cc} &=&
    g (E^x , f_x K_\varphi - \tilde{\mu}_\varphi) f_1 (E^x , f_x K_\varphi - \tilde{\mu}_\varphi)
    \,,\\
    g_{\rm cc} f_2^{\rm cc} &=&
    g (E^x , f_x K_\varphi - \tilde{\mu}_\varphi) f_x \Bigg( 
    f_2 (E^x , f_x K_\varphi - \tilde{\mu}_\varphi)
    \nonumber\\
    &&- K_\varphi \left( \frac{\partial \ln f_x}{\partial E^x}
    - \frac{1}{f_x} \frac{\partial \tilde{\mu}_\varphi}{\partial E^x} \right) \frac{1}{2} \frac{\partial \ln g (E^x , f_x K_\varphi - \tilde{\mu}_\varphi)}{\partial K_\varphi}
    - \left( \frac{\partial \ln f_x}{\partial E^x}
    - \frac{1}{f_x} \frac{\partial \tilde{\mu}_\varphi}{\partial E^x} \right)
    \Bigg)
    \,.
\end{eqnarray}

\subsection{Classical limit}

When solving the anomaly-freedom and covariance equations as outlined above
with $C_{x \varphi}=0$ we should keep in mind that the classical constraint,
(\ref{eq:Hamiltonian constraint - spherical symmetry - Gauss solved}), must be
recovered in an appropriate limit.  The general solution to
(\ref{eq:Covariance equation - (epsilon^0)' term - ((E^x)')^3 term -
  Generalized vacuum - Extended}) is given by
\begin{equation}
    g = \lambda_0 \cos^2 \left( \lambda \left( K_\varphi + \mu_\varphi \right) \right)
    \,,
    \label{eq:Covariant global factor - spherical - generalized vacuum}
\end{equation}
where $\lambda_0$, $\lambda$, and $\mu_\varphi$ are free functions of
$E^x$, and its classical limit is $g \to 1$ as $\lambda_0 \to 1$,
$\lambda \to 0$.
Using this, the general solution to (\ref{eq:Covariance equation - (epsilon^0)' term - (E^x)' term - Generalized vacuum - Extended}), compatible with the classical limit, is
\begin{equation}
    g f_1 = 4 \lambda_0 \left(c_f \frac{\sin (2 \lambda (K_\varphi + \mu_\varphi))}{2 \lambda}
    + q \cos(2 \lambda (K_\varphi + \mu_\varphi))\right)
    \,,
\end{equation}
where $c_f$ and $q$ are free functions of $E^x$, and its classical
limit is $f_1 \to 4 K_\varphi$ as $\lambda_0 , c_f \to 1$ and
$\lambda , \mu_\varphi, q \to 0$.  The general solution
to (\ref{eq:Anomaly-freedom equation (c) updated by covariance (epsilon^0)'''
  term - K_varphi'(E^x)' term - Generalized vacuum - Extended}), compatible
with the classical limit, is then
\begin{equation} \label{f2}
    f_2 =
    - \frac{\alpha_2}{4 E^x}
    + \frac{\sin \left(2 \lambda ( K_{\varphi} + \mu_\varphi) \right)}{2 \lambda \cos^2\left(\lambda (K_{\varphi}+ \mu_\varphi) \right)} \left( \lambda \frac{\partial (\lambda \mu_\varphi)}{\partial E^x}
    + \lambda K_{\varphi} \frac{\partial \lambda}{\partial E^x}\right)
\end{equation}
where $\alpha_2 = \alpha_2 (E^x)$, and its classical limit is
$f_2 \to - 1 / (4 E^x)$ as $\alpha_2\to 1$.
The general
solution to (\ref{eq:Anomlay-free, covariant extra coefficient - Generalized
  vacuum - Extended}) equals
\begin{eqnarray}
    g f_0 &=&
    \lambda_0 \Bigg(
    c_{f 0}
    + \frac{\alpha_0}{E^x}
    + 2 \frac{\sin^2 \left(\lambda  \left(K_{\varphi }+\mu_\varphi \right)\right)}{\lambda^2}\frac{\partial c_{f}}{\partial E^x}
    + 4 \frac{\sin \left(2 \lambda  \left(K_{\varphi }+\mu_\varphi \right)\right)}{2 \lambda} \frac{\partial q}{\partial E^x}
    \nonumber\\
    &&+ 4 c_f \left( \frac{1}{\lambda} \frac{\partial (\lambda \mu_\varphi)}{\partial E^x} \frac{\sin \left(2 \lambda  \left(K_{\varphi }+\mu_\varphi \right)\right)}{2 \lambda}
    + \left(\frac{\alpha_2}{4 E^x} - \frac{\partial \ln \lambda}{\partial E^x}\right) \frac{\sin^2 \left(\lambda  \left(K_{\varphi }+\mu_\varphi \right)\right)}{\lambda^2}\right)
    \nonumber\\
    &&
    + 8 q \left( - \lambda \frac{\partial (\lambda \mu_\varphi)}{\partial E^x} \frac{\sin^2 \left( \lambda \left(K_{\varphi }+\mu_\varphi \right)\right)}{\lambda^2}
    + \left(\frac{\alpha_2}{4 E^x} - \frac{1}{2} \frac{\partial \ln \lambda}{\partial E^x}\right) \frac{\sin \left(2 \lambda  \left(K_{\varphi }+\mu_\varphi \right)\right)}{2 \lambda} \right)
    \nonumber\\
    &&+ 4 K_{\varphi} \frac{\partial \ln \lambda}{\partial E^x} \left( c_f \frac{\sin \left(2 \lambda  \left(K_{\varphi }+\mu _{\varphi }\right)\right)}{2 \lambda}
    + q \cos \left(2 \lambda  \left(K_{\varphi }+\mu _{\varphi }\right)\right) \right)
    \Bigg) \label{f0}
\end{eqnarray}
where $c_{f 0}$ and $\alpha_0$ are undetermined functions of $E^x$. (They can
be combined to a single free function, but it is convenient to separate them
for the purpose of taking the classical limit.)  Its classical limit is
$g f_0 \to 1 / E^x$ as $\alpha_0, \alpha_2, \lambda_0 , c_{f} \to 1$,
$\lambda , q , \mu_\varphi$.

These results completely determine the anomaly-free, covariant Hamiltonian
constraint and the structure function for vacuum up to the undetermined
functions of $E^x$.  The classical constraint (\ref{eq:Hamiltonian
  constraint - spherical symmetry - Gauss solved}) can be recovered in
different limits. The most straightforward way is to set all the parameters
constant and then take the limits $\lambda_0 , c_{f} , \alpha_i \to 1$ and
$\lambda, c_{f 0}, c_{f 2}, \mu_\varphi \to 0$. The cosmological constant can
be recovered by instead setting $c_{f0} \to - \Lambda$.

\subsection{Periodicity and bounded-curvature effects}

The additional restrictions on consistent modified Hamiltonian constraints,
implied by the covariance condition, allow us to clarify the question of
possible functional dependences of the constraint on $K_{\varphi}$, with
various properties of relevance for models of loop quantum gravity.

The parameter $\lambda$ can heuristically be interpreted as the holonomy
angular length in models of loop quantum gravity. We may therefore restrict
its form by referring to specific triangulations of space.  For example, we
can choose a fine lattice such that the spheres are triangulated by small
squares of side length $\lambda$.  Each plaquette at radius $\sqrt{E^x}$ then
covers an area $E^x \lambda^2$.  Requiring that the plaquettes at different
radii have equal sizes, we obtain
\begin{equation}
    \lambda = \frac{\bar{r}}{\sqrt{E^x}} \bar{\lambda}
\end{equation}
where $\bar{r}$ is a constant reference radius at which
$\lambda = \bar{\lambda}$.  This result satisfies $\lambda \to 0$ as
$E^x \to \infty$, as desired to recover the classical limit at large
distances. 

Furthermore, the Hamiltonian constraint obtained from the previous results
would be non-periodic in $K_\varphi$ for non-constant $\lambda$, owing to the
last term in (\ref{f2}) and (\ref{f0}).  In models of loop quantum gravity, a Hamiltonian
periodic in $K_{\varphi}$ (if not $K_x$, which is harder to achieve) is often
desired in order to motivate a well-defined quantization in a representation
of the holonomy-flux algebra. In addition to being quite restricted by the
covariance condition, such periodicity properties are not invariant under
canonical transformations.  The canonical transformations
(\ref{eq:Diffeomorphism-constraint-preserving canonical transformations of
  Constraint terms - Spherical - residual after modulo}) can be used to
re-establish periodicity if we start with a non-constant $\lambda$ for which
the last term in (\ref{f2}) and (\ref{f0}) is not periodic.

Upon such a transformation with $\tilde{\mu}_\varphi=\mu_\varphi$, the
Hamiltonian constraint terms become
\begin{eqnarray}
    g_{\rm cc}
    &=&
    \lambda_0 f_x \cos^2 \left( \lambda f_x K_\varphi \right)\\
    g_{\rm c c} f_0^{\rm cc}
    &=& \frac{\lambda_0}{f_x} \Bigg( c_{f 0}
    + \frac{\alpha_0}{E^x}
    + 2 \frac{\sin^2 \left(\lambda f_x K_\varphi\right)}{\lambda^2}\frac{\partial c_{f}}{\partial E^x}
    + 4 \frac{\sin \left(2 \lambda f_x K_\varphi\right)}{2 \lambda} \frac{\partial q}{\partial E^x}
    \nonumber\\
    &&+ 4 c_f \left(\frac{\alpha_2}{4 E^x} - \frac{\partial \ln \lambda}{\partial E^x}\right) \frac{\sin^2 \left(\lambda f_x K_\varphi\right)}{\lambda^2}
    \nonumber\\
    &&
    + 8 q \left(\frac{\alpha_2}{4 E^x} - \frac{1}{2} \frac{\partial \ln \lambda}{\partial E^x}\right) \frac{\sin \left(2 \lambda f_x K_\varphi\right)}{2 \lambda}
    \Bigg)
    \nonumber\\
    &&+ \frac{\partial \ln (\lambda f_x )}{\partial E^x} 4 \lambda_0 K_\varphi \left(c_f \frac{\sin (2 \lambda f_x K_\varphi)}{2 \lambda}
    + q \cos(2 \lambda f_x K_\varphi)\right)\\
    g_{\rm cc} f_1^{\rm cc} &=&
    4 \lambda_0 \left(c_f \frac{\sin (2 \lambda f_x K_\varphi)}{2 \lambda}
    + q \cos(2 \lambda f_x K_\varphi)\right)\\
    g_{\rm cc} f_2^{\rm cc}
    &=& \lambda_0 f_x \cos^2 \left( \lambda f_x K_\varphi \right) \Bigg(
    - \frac{\alpha_2}{4 E^x}
    - \frac{\partial \ln f_x}{\partial E^x}
    \nonumber\\
    &&
    + \frac{\sin \left(2 \lambda f_x K_\varphi \right)}{2 \lambda \cos^2\left(\lambda f_x K_\varphi \right)} \lambda^2 f_x K_\varphi \frac{\partial \ln (\lambda f_x)}{\partial E^x}
    \Bigg)
    \,.
\end{eqnarray}
The phase $\mu_\varphi$ then drops out of the Hamiltonian constraint.
Furthermore, the constraint is rendered periodic and bounded in $K_\varphi$ by
choosing
\begin{equation}
    f_x = \frac{\bar{\lambda}}{\lambda}
\end{equation}
where $\bar{\lambda}$ is a constant. However, this transformation also removes
any $E^x$-dependence in coefficients of $K_{\varphi}$ in periodic functions.

Building upon the constraint obtained by the preceding canonical
transformation, a further simplification consists in redefining the remaining
parameters according to
\begin{equation}
    \lambda_0 \to \lambda_0 \frac{\lambda}{\bar{\lambda}}
    \ , \
    q \to q \frac{\bar{\lambda}}{\lambda}
    \ , \
    c_{f 0} \to \frac{\bar{\lambda}^2}{\lambda^2} c_{f 0}
    \ , \
    \alpha_0 \to \frac{\bar{\lambda}^2}{\lambda^2} \alpha_0
    \ , \
    \alpha_2 \to \alpha_2 + 4 E^x \frac{\partial \ln \lambda}{\partial E^x}
    \,.
    \label{eq:Redefinitions of lambda}
\end{equation}
Under these redefinitions, the most general Hamiltonian constraint, up to
canonical transformations, is of the form
\begin{eqnarray}
    H
    &=& - \lambda_0 \frac{\sqrt{E^x}}{2} \Bigg[ E^\varphi \Bigg(
    c_{f 0}
    + \frac{\alpha_0}{E^x}
    + 2 \frac{\sin^2 \left(\bar{\lambda} K_\varphi\right)}{\bar{\lambda}^2}\frac{\partial c_{f}}{\partial E^x}
    + 4 \frac{\sin \left(2 \bar{\lambda} K_\varphi\right)}{2 \bar{\lambda}} \frac{\partial q}{\partial E^x}
    \nonumber\\
    &&+ 4 c_f \frac{\alpha_2}{4 E^x} \frac{\sin^2 \left(\bar{\lambda} K_\varphi\right)}{\bar{\lambda}^2}
    + 8 q \frac{\alpha_2}{4 E^x} \frac{\sin \left(2 \bar{\lambda} K_\varphi\right)}{2 \bar{\lambda}}
    \Bigg)
    + 4 K_x \left(c_f \frac{\sin (2 \bar{\lambda} K_\varphi)}{2 \bar{\lambda}}
    + q \cos(2 \bar{\lambda} K_\varphi)\right)
    \nonumber\\
    &&+ \frac{((E^x)')^2}{E^\varphi} \left(
    - \frac{\alpha_2}{4 E^x} \cos^2 \left( \bar{\lambda} K_\varphi \right)
    + \frac{K_x}{E^\varphi} \bar{\lambda}^2 \frac{\sin \left(2 \bar{\lambda} K_\varphi \right)}{2 \bar{\lambda}}
    \right)
    \nonumber\\
    &&
    + \left( \frac{(E^x)' (E^\varphi)'}{(E^\varphi)^2}
    - \frac{(E^x)''}{E^\varphi} \right) \cos^2 \left( \bar{\lambda} K_\varphi \right)
    \Bigg]
    \nonumber\\
    &=& - \lambda_0 \frac{\sqrt{E^x}}{2} E^\varphi \Bigg[
    c_{f 0}
    + \frac{\alpha_0}{E^x}
    + 2 \frac{\sin^2 \left(\bar{\lambda} K_\varphi\right)}{\bar{\lambda}^2}\frac{\partial c_{f}}{\partial E^x}
    + 4 \frac{\sin \left(2 \bar{\lambda} K_\varphi\right)}{2 \bar{\lambda}} \frac{\partial q}{\partial E^x}
    + \frac{\alpha_2}{\bar{\lambda}^2 E^x} c_f
    \nonumber\\
    &&- \frac{(E^\varphi)^2}{\lambda_0^2 \bar{\lambda}^2 E^x} \left( \frac{\alpha_2}{E^x} \tilde{q}^{x x}
    + 2 \frac{K_x}{E^\varphi} \frac{\partial \tilde{q}^{x x}}{\partial K_\varphi} \right)
    - \left( \frac{(E^x)' \left((E^\varphi)^{- 2}\right)'}{2}
    + \frac{(E^x)''}{(E^\varphi)^2} \right) \cos^2 \left( \bar{\lambda} K_\varphi \right)
    \Bigg]
    \label{eq:Hamiltonian constraint - Anomaly-free - covariant - spherical - vacuum - general - Fully canonically moded - NO lambda}
\end{eqnarray}
with structure function
\begin{equation}
    \tilde{q}^{x x}
    =
    \left(
    \left( c_{f}
    + \left(\frac{\bar{\lambda} (E^x)'}{2 E^\varphi} \right)^2 \right) \cos^2 \left(\bar{\lambda} K_\varphi\right)
    - 2 q \bar{\lambda}^2 \frac{\sin \left(2 \bar{\lambda} K_\varphi\right)}{2 \bar{\lambda}}\right)
    \lambda_0^2 \frac{E^x}{(E^\varphi)^2}
    \,.
    \label{eq:Structure function - Anomaly-free - covariant - spherical - vacuum - general - Fully canonically moded - NO lambda}
\end{equation}
The function $\lambda$ has been completely absorbed by the other parameters.
Thus, we conclude that non-constant $\lambda$ can always be traded in for
non-classical functions for the other parameters, but not in an obvious
way. Within the setting of modified gravity, there is no invariant meaning to
specific $E^x$-dependencies or periodicity conditions in holonomies. The
classical constraint is recovered in the limit
$\lambda_0 , c_{f} , \alpha_i \to 1$ and
$\bar{\lambda}, c_{f 0}, c_{f 2} \to 0$. The cosmological constant can be
recovered by instead setting $c_{f0} \to - \Lambda$.  In the following
sections we drop the bar in $\bar{\lambda}$ and write this \emph{constant}
parameter as $\lambda$ for simplicity.

The case $\lambda_0 , c_{f} , \alpha_i \to 1$ and $c_{f 0}, q \to 0$ for
(\ref{eq:Hamiltonian constraint - Anomaly-free - covariant - spherical -
  vacuum - general - Fully canonically moded - NO lambda}) was first found in
\cite{SphSymmMatter} by demanding anomaly-freedom and, since they had no
knowledge of the covariance condition derived here, some functions were only
proposed and the rest obtained by solving the anomaly-freedom equations for a
less general constraint than (\ref{eq:Hamiltonian constraint ansatz -
  Generalized vacuum - Extended}).  They also chose constant $\lambda$ in
order to have a constraint periodic in $K_{\varphi}$.

In \cite{SphSymmMatter}, it was shown that the Hamiltonian constraint in this
case is the result of a specific linear combination of the classical
constraints with phase-space dependent coefficients, after performing the
diffeomorphism-constraint preserving canonical transformation
\begin{equation}
    K_\varphi \to \frac{\sin (\lambda K_\varphi)}{\lambda}
    \quad,\quad
    E^\varphi \to \frac{E^\varphi}{\cos (\lambda K_\varphi)}
    \,.
\end{equation}
The emergent space-time it implies was studied in \cite{SphSymmEff}, where it
was shown that the classical singularity does not appear in a black hole-like
solution. Our analysis demonstrates that this outcome is an implication of the
phase-space dependent linear combination, rather than of periodic and bounded
functions in the Hamiltonian constraint. Our derivations strengthen this
result by showing in Section~\ref{sec:Lin} that this constraint is the unique
covariant linear combination of the constraints up to an overall function of
$E^x$.

In the constraint (\ref{eq:Hamiltonian constraint - Anomaly-free - covariant -
  spherical - vacuum - general - Fully canonically moded - NO lambda}), the
terms containing $q$ are the only modifications with non-trivial holonomy
effects allowed by anomaly-freedom and covariance that have not previously
been considered. However, they do not seem directly related to holonomy
corrections in any obvious way as some of them survive the limit
$\lambda \to 0$. Among all the modification functions found here, $c_f$,
$\lambda$, $q$ and $\lambda_0$ are the most characteristic of emergent
modified gravity: If we require the classical form of the structure function,
these functions must all take their classical expressions. The freedom of choosing
the remaining functions is related to the emergent metric only through their
dependence on a non-constant $\lambda$, but the freedom expressed by these
functions appears even classically in the terms of a
$1+1$-dimensional dilaton action. The same argument shows that the constraints
depend at most quadratically on $K_{\varphi}$ unless a non-classical emergent
metric is considered, which requires $\lambda\not=0$.

\subsection{General partial Abelianization}
\label{sec:PartialGen}

In a combination of our preceding results, we can now use linear combinations
of the form $H^{({\rm new})} = B H + A H_r$, where $A$ and $B$ are phase-space
dependent functions as in Section~\ref{sec:Lin}, but insert the general modified
Hamiltonian constraint $H$ from the present section, given in equation
(\ref{eq:Hamiltonian constraint - Anomaly-free - covariant - spherical -
  vacuum - general - Fully canonically moded - NO lambda}). Since the
Hamiltonian constraint is general, this procedure should not result in new
covariant theories, but we can use the construction as in
Section~\ref{sec:Partial} and impose conditions other than general covariance
on the resulting structure function $\tilde{q}^{xx}_{({\rm new})}$. As an example, we derive
new partial Abelianizations by requiring that the new structure function
vanish. The resulting theory would be compatible with general covariance
because this condition has been implemented on the original Hamiltonian
constraint, and it might be more amenable to quantizations using, for
instance, the loop representation because structure functions have been
eliminated in the partial Abelianization.

Assuming some function $B = B (E^x , K_\varphi )$, the
second function $A$ in the linear combination and the structure function are
uniquely determined, now given by
\begin{equation}
    A     =
    - \lambda_0 \cos^2 (\bar{\lambda} K_\varphi) \frac{\sqrt{E^x} (E^x)'}{2 (E^\varphi)^2} \frac{\partial B}{\partial K_\varphi}
    \ ,
    \label{eq:A coefficient - Linear combination General}
\end{equation}
and
\begin{eqnarray}
    \tilde{q}^{x x}_{({\rm new})}
    &=&
    B \frac{E^x}{(E^\varphi)^2} \frac{\lambda_0 \cos (\lambda K_\varphi)}{2}
    \Bigg(
    2 \lambda_0 B \left( c_{f} \cos (\bar{\lambda} K_\varphi)
    - 2 \lambda^2  q \frac{\sin (\bar{\lambda} K_\varphi)}{\bar{\lambda}}\right)
   \label{eq:Structure function - Linear combination General}  \\
    &&- 2 \lambda_0 \frac{\partial B}{\partial K_\varphi} \cos \left(\bar{\lambda} K_\varphi\right) \left( c_{f} \frac{\sin \left(2 \bar{\lambda} K_\varphi\right)}{2 \bar{\lambda}}
    + \frac{q}{2} \left( \cos \left( \lambda K_\varphi\right) + \cos \left(3 \lambda K_\varphi\right) \right)\right)
    \nonumber\\
    &&
    + \frac{\left(\left(E^x\right)'\right)^2}{(E^\varphi)^2} \lambda_0 \cos (\bar{\lambda} K_\varphi)
    \bigg( B \bar{\lambda}^2
    - \frac{\partial B}{\partial K_\varphi} 3 \bar{\lambda}^2 \frac{\sin \left(2 \bar{\lambda} K_\varphi\right)}{2 \bar{\lambda}}
    + \frac{\partial^2 B}{(\partial K_\varphi)^2} \cos^2 (\bar{\lambda} K_\varphi) \bigg)
    \Bigg)\nonumber
\end{eqnarray}
if we apply the procedure to the general expression of a modified Hamiltonian constraint.

Partial Abelianization requires $\tilde{q}^{x x}_{({\rm new})}=0$. Because $B$
is assumed to be independent of $(E^x)'$, the first two lines of
(\ref{eq:Structure function - Linear combination}) must vanish separately,
such that
\begin{equation}
    B= \frac{B_x}{\cos^2 (\bar{\lambda} K_\varphi)} \left(c_{f} \frac{\sin \left(2 \bar{\lambda} K_\varphi\right)}{2 \bar{\lambda}}
    + q \cos \left(2 \bar{\lambda} K_\varphi\right)\right)
\end{equation}
where $B_x = B_x (E^x)$.
Inserting  this result in the last two lines, the condition that they vanish too implies
\begin{equation}
    B_x \bar{\lambda} q = 0
    \,.
\end{equation}
For a non-trivial Abelianization with $\bar{\lambda} \neq 0$, this is realized
only if $q = 0$. We arrive at the Abelianized constraint
\begin{eqnarray}
    \frac{H^{({\rm A})}}{B_x} &=&
    - \frac{\sqrt{E^x}}{2} \frac{\tan (\bar{\lambda} K_\varphi)}{\bar{\lambda}} \Bigg[
    E^\varphi \Bigg(
    c_{f 0}
    + \frac{\alpha_0}{E^x}
    + 2 \frac{\sin^2 \left(\bar{\lambda} K_\varphi\right)}{\bar{\lambda}^2}\frac{\partial c_{f}}{\partial E^x}
    + c_f \frac{\alpha_2}{E^x} \frac{\sin^2 \left(\bar{\lambda} K_\varphi \right)}{\bar{\lambda}^2}
    \Bigg)
    \nonumber\\
    &&+ 4 K_x c_f \frac{\sin (2 \bar{\lambda} K_\varphi)}{2 \bar{\lambda}}
    + \frac{((E^x)')^2}{E^\varphi} \left(
    - \frac{\alpha_2}{4 E^x} \cos^2 \left( \bar{\lambda} K_\varphi \right)
    + \frac{K_x}{E^\varphi} 2 \bar{\lambda}^2 \frac{\sin \left(2 \bar{\lambda} K_\varphi \right)}{2 \bar{\lambda}}
    \right)
    \nonumber\\
    &&
    + \left( \frac{(E^x)' (E^\varphi)'}{(E^\varphi)^2}
    - \frac{(E^x)''}{E^\varphi} \right) \cos^2 \left( \bar{\lambda} K_\varphi \right)
    \Bigg]
    - \frac{\sqrt{E^x} \left(E^x\right)'}{2 (E^\varphi)^2} \left( E^\varphi K_\varphi' - K_x (E^x)' \right)
    \ ,
    \label{eq:Covariant Hamiltonian constraint - spherical symmetry - vacuum - general - Abelian}
\end{eqnarray}
where we redefined $B_x$ in order to absorb an overall term $c_{f} \lambda_0$.
This result agrees with the partial Abelianization obtained in
Section~\ref{sec:Partial} if we choose all parameters except for
$\bar{\lambda}$ to take classical values.

The Abelianized constraint $H^{({\rm A})}$ has a divergence at
$\bar{\lambda} K_\varphi = \pi / 2$ implied by the overall
$\tan (\bar{\lambda} K_\varphi)$ multiplying the terms of the first line.
Looking at the first three terms, the divergence can be resolved by setting
\begin{equation}
    \frac{\partial c_f}{\partial E^x} = - \frac{\bar{\lambda}^2}{2} \left( c_{f 0} +\frac{\alpha_0}{E^x} \right)
    \ ,
    \label{eq:alpha_0 - Partial Abelianization condition}
\end{equation}
which is easily solved for $c_f$ if we use the classical values $\alpha_0 = 1$
and $c_{f0}=-\Lambda$: 
\begin{equation}
    c_f = 1 + \frac{\bar{\lambda}^2}{2} ( \Lambda E^x - \ln (E^x / c_0))
    \ ,
    \label{eq:c_f - simple Partial Abelianization condition}
\end{equation}
where $c_0$ is the integration constant. This non-classical form of the
function $c_f$, depending on $\bar{\lambda}$, may be considered an indirect
effect of non-constant holonomy length or a modified angular metric. (Trying
to include the divergence of the last term of the first line in
(\ref{eq:Covariant Hamiltonian constraint - spherical symmetry - vacuum -
  general - Abelian}) in this derivation does not result in a $c_f$ compatible
with the classical limit for $\bar{\lambda}\to0$.)

The first three terms of the first line are then free of divergences in $K_\varphi$
while respecting the classical limit of the Hamiltonian constraint
(\ref{eq:Hamiltonian constraint - Anomaly-free - covariant - spherical -
  vacuum - general - Fully canonically moded - NO lambda}), a feat that could
not be accomplished in Section~\ref{sec:Partial} because
non-classical forms of $c_f$ were not considered there.
However, divergence of the last term in the first line remains unresolved.

The last line does not show an immediate divergence in $K_\varphi$, but there
is one if we 
put $K_\varphi'$ into a manifestly periodic form:
\begin{equation}
   K_\varphi' =
\sec ( \beta \bar{\lambda} K_\varphi ) \frac{\sin (
    \beta \bar{\lambda} K_\varphi )'}{\beta \bar{\lambda}}
\end{equation}
with some integer $\beta$.  The appearance of
$\sec (\beta \bar{\lambda} K_\varphi)$ makes it difficult to promote
$H^{({\rm A})}_{\rm c c}$ to an operator in a loop quantization because of its
divergence at $\beta \bar{\lambda} K_\varphi = \pi / 2$.  One possible
resolution might be to consider a modified diffeomorphism constraint as in
\cite{DiffeoOp,VectorHol}, since $K_{\varphi}'$ is introduced by taking a
linear combination with this constraint. This divergence problem is therefore
related to the fact \cite{ALMMT} that the diffeomorphism constraint, as the
generator of infinitesimal diffeomorphisms, cannot be directly quantized in
the usual loop representation but is replaced by the action of finite
diffeomorphisms. We leave this problem for future work.

A notable difference between our result (\ref{eq:Covariant Hamiltonian
  constraint - spherical symmetry - vacuum - general - Abelian}) and the
Abelianization of \cite{LoopSchwarz,LoopSchwarz2} is that the latter does not
depend on $K_x$, while the former does. This term is harder to express in a
loop representation because so far no consistent modification periodic in
$K_x$ has been found, but it demonstrates that our result is much more general
than the previous partial Abelianizations. If the $K_x$-dependence is
completely eliminated, while spatial derivatives of $E^{\varphi}$ are also
removed by the construction of \cite{LoopSchwarz}, the constraints trivially
Poisson commute. However, removing derivatives of $E^{\varphi}$ requires
integrating by parts, which introduces a certain degree of non-locality. In
our constraint, neither $K_x$ not $(E^{\varphi})'$ have been eliminated, and
we only used linear combinations without integrating by parts. Our
construction is completely local and relies on highly non-trivial
cancellations of several terms for the new constraints to Poisson commute.

\section{Conclusions}

Our analysis of canonical gauge transformations acting on a space-time metric
has revealed gaps in the widely held assumption that the constraint brackets
in canonical models of modified gravity have full control over general
covariance.  These brackets ensure the correct transformation of the ``time''
components of a compatible space-time metric, as previously recognized, but by
themselves they do not guarantee the correct transformation of the spatial
metric, determined by the structure function in hypersurface-deformation
brackets, to reproduce full space-time diffeomorphisms on shell.  The new
covariance condition formulated here is automatically satisfied if the
structure function depends directly on a single phase-space variable, as in
the classical case where the structure function is the inverse spatial metric.
But this is no longer the case if the structure function is a composite field
and depends on multiple phase-space functions, for instance in modified
theories in which it may also depend on momentum components.  The full
covariance condition is essential in canonical theories of modified gravity,
where it presents strong restrictions on the allowed modifications.

We have specifically applied the covariance condition to the spherically
symmetric model in which the classical constraints are replaced by phase-space
dependent linear combinations of the gauge generators. As we discussed in the
introduction, modifications are possible because such linear combinations in
the context of hypersurface deformations imply a redefinition of the normal
direction. The normal, together with the spatial metric derived from the
structure function of the constraint brackets, then determines an emergent
space-time metric which need not be equivalent to the original classical
geometry.  Our explicit derivations in the spherically symmetric model, where
the relevant equations that control general covariance can be solved exactly,
confirm this expectation. We derived a new covariant model in which signature
change may be possible, and confirmed the covariance of a recent model derived
initially by different means \cite{SphSymmEff}. Our results also demonstrate
the covariance of older models, such as \cite{LTBII}.  Other examples, such as
\cite{LoopSchwarz2,EffLine}, turned out not to be covariant.

In this process, we have developed a general method to obtain anomaly-free
brackets from the linear combination of some of the original constraints that
resulted in the computation of a new emergent space-time and a well-defined,
off-shell partial Abelianization along the lines of
\cite{LoopSchwarz,LoopSchwarz2}.  This result opens the way to analyzing more
complicated modified constraints and their emergent space-times, and it
restricts the modifications to those compatible with general covariance, which
had proved challenging until now.

Finally, we have derived a general expression for modified Hamiltonian
constraints compatible with general covariance, extending the vacuum results of
\cite{SphSymmMatter,SphSymmMatter2} by implementing the latter condition. A
discussion of canonical transformations implied several simplifications
and revealed redundancies in common choices of modifications in models of loop
quantum gravity, in particular in the choice of periodic functions with
phase-space dependent periods. As a by-product, we derived new non-trivial
partial Abelianizations of constraints compatible with general covariance.

\section*{Acknowledgements}

We are grateful to Idrus Belfaqih, Suddho Brahma and Simone Speziale for discussions.  Our
presentation benefited from MB's participation in the ESI workshop ``Higher
structures and field theory.'' Support from the Erwin Schr\"odinger Institute
at the University of Vienna is gratefully acknowledged. This work was
supported in part by NSF grant PHY-2206591.

\begin{appendix}

\section{Covariance of the emergent extrinsic-curvature tensor}
\label{a:ExtCurv}

Extrinsic curvature is defined as a Lie derivative of the spatial metric along
the unit normal to a hypersurface. An analysis of how the tensor transforms therefore
requires an equation for the transformation of the unit normal by a gauge
transformation that changes the space-time slicing. In this appendix, we
assume that there is a covariant space-time line element with space-time
metric $g_{\mu\nu}$. In our canonical theories, $g_{\mu\nu}$ would be the
emergent metric tensor, but here we drop tildes for the sake of convenience.

Starting with
  \begin{equation}
    q_{\mu \nu} = g_{\mu \nu} + n_\mu n_\nu
    \ ,
\end{equation}
both the unit normal vector and the spatial metric change to
$n^\mu + \delta_\epsilon n^\mu$ and $q_{a b} + \delta q_{a b}$ under a gauge
transformation.  Since $n^{\mu}$ is assumed to be normalized, we have
$n^\mu q_{\mu \nu} = 0$. It will remain normalalized after the gauge
transformation if and only if $\delta_\epsilon (N n^\mu q_{\mu \nu}) = 0$.

It is easier to evaluate this condition if we use the space-time metric
instead of the spatial metric, in which case we can express normalization of
$n^{\mu}$ as $N n^\mu g_{\mu \nu} {\rm d} x^\nu = \sigma N^2 {\rm d} t$ where
$\sigma=1$ for Lorentzian signature and $\sigma=-1$ for Euclidean signature.
Normalization is then preserved by a gauge transformation $\delta_{\epsilon}$
if and only if
\begin{equation}
    \delta_\epsilon (N n^\mu g_{\mu \nu}) d x^\nu
    = 2 \sigma N \delta_\epsilon N d t
    \,,
    \label{eq:Unit normal condition - Geometrodynamics}
\end{equation}
where we shall use the transformation of the lapse (\ref{eq:Off-shell gauge
  transformations for lapse and shift}).  We can now expand the left-hand side
using the Leibniz rule to obtain
$n^\mu \delta_\epsilon (N g_{\mu \nu}) + N g_{\mu \nu} \delta_\epsilon n^\mu$.
The change of basis $n^\mu = N^{-1} (t^\mu-N^a s^\mu_a)$ and the component
expression $g^{t \mu}$ of in terms of the lapse function and shift vector then
imply
\begin{equation}
    \delta_\epsilon n^\mu =
    - \frac{1}{N} \delta_\epsilon N n^\mu
    - \frac{1}{N} \delta_\epsilon N^a s^{\mu}_a
    \,.
    \label{eq:Unit normal vector gauge transformation}
\end{equation}

The normal vector is associated to the particular coordinates and foliation we
choose, therefore, its transformation is not directly equivalent to a Lie
derivative. (This is similar to how connections do not transform by a simple
Lie derivative.)  In fact, the Lie derivative of $n^{\mu}$ and its infinitesimal coordinate
transformation are related by
\begin{eqnarray}
    \mathcal{L}_\xi n^\mu &=&
    \xi^\nu \partial_\nu n^\mu
    - n^\nu \partial_\nu \xi^\mu
    \nonumber\\
    &=&
    \frac{\epsilon^0}{N} \left( \partial_t \left(\frac{1}{N}\right) t^\mu - \partial_t \left(\frac{N^a}{N}\right) s_a^\mu \right)
    + \left( \epsilon^b - \frac{\epsilon^0}{N} N^b \right) \left( \partial_b \left(\frac{1}{N}\right) t^\mu - \partial_b \left(\frac{N^a}{N}\right) s_a^\mu \right)
    \nonumber\\
    &&
    - \frac{1}{N} \left( \partial_t \left(\frac{\epsilon^0}{N}\right) t^\mu + \left( \partial_t \epsilon^a - \partial_t \left(\frac{\epsilon^0}{N} N^a \right)\right) s_a^\mu \right)
    \nonumber\\
    &&
    + \frac{N^b}{N} \left( \partial_b \left(\frac{\epsilon^0}{N}\right) t^\mu + \left( \partial_b \epsilon^a - \partial_b \left(\frac{\epsilon^0}{N} N^a\right) \right) s_a^\mu \right)
    \nonumber\\
    &=&
    - \frac{1}{N} \left( \Dot{\epsilon}^0
    + \partial_b N
    - N^b \partial_b \epsilon^0 \right) n^\mu
    - \frac{1}{N} \left( \Dot{\epsilon}^a
    + \epsilon^b \partial_b N^a
    - N^b \partial_b \epsilon^a \right) s_a^\mu
    \nonumber\\
    &=&
    - \frac{\delta_\epsilon N}{N} n^\mu
    - \frac{\delta_\epsilon N^a - q^{a b} \left(\epsilon^0 \partial_b N - N \partial_b \epsilon^0 \right)}{N} s_a^\mu
    \nonumber\\
    &=&
    \delta_\epsilon n^\mu + \frac{q^{a b}}{N} \left(\epsilon^0 \partial_b N - N \partial_b \epsilon^0 \right) s_a^\mu
    \,.
\end{eqnarray}

In addition to the normal vector, the spatial basis vectors $s_a^{\mu}$ appear
in some of the expressions. Before we apply our results to extrinsic
curvature, we make sure that $s_a^{\mu}$ does not change by a gauge
transformation.  For these vectors to remain spatial, we have $\delta_\epsilon
(g_{\mu \nu} n^\mu s^\nu_b) = 0$. We evaluate this condition by decomposing
\begin{equation}
    \delta_\epsilon s^\mu_b \equiv
    A_b n^\mu
    + B s^\mu_b
\end{equation}
into normal and spatial components. Using (\ref{eq:Unit normal vector gauge
  transformation}) and the gauge transformations of lapse and shift, the
condition that $s_a^{\mu}$ remain spatial directly implies
\begin{equation}
    A_b =
    g_{\mu b} \delta_\epsilon n^\mu
    + n^\mu \delta_\epsilon g_{\mu b}
    = 0
    \,.
\end{equation}
Furthermore,
$\delta_\epsilon q_{a b} = \delta_\epsilon (g_{\mu \nu} s^\mu_a s^\nu_b)$
implies $B=0$.

Now, the extrinsic-curvature tensor is given by
\begin{equation}
    K_{\mu \nu} = \frac{1}{2} \mathcal{L}_n q_{\mu \nu}
\end{equation}
and therefore depends on the slicing through the unit normal. Its
transformation should reflect the slicing dependence because $n^{\mu}$ changes
by a gauge transformation, as we saw.  If we consider the infinitesimal
changes
\begin{eqnarray}
    q_{\mu \nu} &\to& q_{\mu \nu} + \mathcal{L}_\xi q_{\mu \nu}
    \ , \\
    n^\mu &\to& n^\mu + \delta_\epsilon n^\mu
\end{eqnarray}
under a coordinate transformation, where $\delta_\epsilon n^\mu$ is given by
(\ref{eq:Unit normal vector gauge transformation}),
 the extrinsic-curvature tensor transforms as
\begin{eqnarray}
    K_{\mu \nu} &\to& \frac{1}{2} \mathcal{L}_{n + \delta_\epsilon n} \left(
                      q_{\mu \nu} + \mathcal{L}_\xi q_{\mu \nu} \right) 
    \nonumber\\
    && = K_{\mu\nu} + \frac{1}{2} \mathcal{L}_{\delta_\epsilon n} q_{\mu \nu}
    + \frac{1}{2} \mathcal{L}_n \mathcal{L}_\xi q_{\mu \nu}
\end{eqnarray}
where we have kept only the first-order term in $\xi^\mu$ for an infinitesimal
transformation.
For a gauge transformation of the same tensor, we have
\begin{eqnarray}
    \delta_\epsilon K_{\mu \nu} &=&
    \delta_\epsilon \left(\frac{1}{2} \mathcal{L}_n q_{\mu \nu} \right)\nonumber\\
    &=& \frac{1}{2} \mathcal{L}_{\delta_\epsilon  n} q_{\mu \nu}
    + \frac{1}{2} \mathcal{L}_n \left( \delta_\epsilon q_{\mu \nu} \right)
    \,.
\end{eqnarray}
If the covariance condition,
$\delta_\epsilon q_{\mu \nu} = \mathcal{L}_\xi q_{\mu \nu}$, is satisfied,
this is precisely the expression derived above. Therefore, the gauge
transformation of extrinsic curvature, derived from the emergent space-time
metric, gives the desired covariant transformation, keeping in mind that the
coordinate transformation is not a Lie derivative which is similar to the
transformation of the normal vector or connections, as all these cases depend
on the foliation. (Extrinsic curvature is a spatial tensor on a given slice
but not a space-time tensor.)

Since the spatial metric and extrinsic curvature form a complete set of
spatial tensors that define the geometry of an embedded hypersurface, we
conclude that the gauge transformation of all tensors derived from the
spacetime metric will be equivalent to their infinitesimal coordinate
transformations provided the covariance condition
$\delta_\epsilon g_{\mu \nu} = \mathcal{L}_\xi g_{\mu \nu}$ is satisfied.

  \section{Restrictions on the general Hamiltonian constraint from $\{H,H_r\}$}
  \label{appHam}
  
The third term in (\ref{HHr}) must vanish, $\mathcal{F}_2 = 0$, which implies
\begin{eqnarray}
    0 &=&
    a_2 + \left(a_{x \varphi} - 3 \frac{\partial d_2}{\partial E^x}\right) E^\varphi
    \ , \\
    0 &=&
    2 a_{\varphi \varphi}
    - 3 \frac{\partial d_2}{\partial E^\varphi}
    \ , \\
    0 &=&
    b_2 + \left(c_{\varphi \varphi} - 3 \frac{\partial d_2}{\partial K_\varphi}\right) E^\varphi
    \ , \\
    0 &=&
    \frac{\partial d_2}{\partial K_x}
    \,.
\end{eqnarray}
Using this and $\mathcal{F}_0 = 0$, we obtain
\begin{equation}
    \frac{\partial}{\partial E^\varphi} \left(
    E^{\varphi} \frac{\partial}{\partial E^\varphi} \left( a_{\varphi \varphi}
    - \frac{\partial d_2}{\partial E^\varphi} \right)
    - K_x \frac{\partial a_{\varphi \varphi}}{\partial K_x} \right)
    = 0
  \end{equation}
  from the coefficient of $(E^\varphi)'''$,
  \begin{equation}
    \frac{\partial}{\partial K_\varphi} \left(
    E^{\varphi} \left(
     \frac{\partial b_{\varphi \varphi}}{\partial E^\varphi}
    - \frac{\partial c_{\varphi \varphi}}{\partial K_\varphi}
    + \frac{\partial^2 d_2}{(\partial K_\varphi)^2} \right)
    + K_x \frac{\partial b_{\varphi \varphi}}{\partial K_x}
    + b_{\varphi \varphi}
    - \frac{\partial b_2}{\partial K_\varphi}
    \right)
    = 0
\end{equation}
from the coefficient of $K_\varphi'''$, and
\begin{equation}
    \frac{\partial}{\partial E^x} \left(
    E^{\varphi} \left( - \frac{\partial a_{x \varphi}}{\partial E^x}
    + \frac{\partial a_{x x}}{\partial E^\varphi}
    + \frac{\partial^2 d_2}{(\partial E^x)^2}\right)
    + K_x \frac{\partial a_{x x}}{\partial K_x}
    + a_{x x}
    - \frac{\partial a_2}{\partial E^x} \right)
    = 0
  \end{equation}
  from the coefficient of $(E^x)'''$.

Using these results and $\mathcal{F}_1 = - H$, we obtain
\begin{equation}
    d_2 + 2 \frac{\partial d_2}{\partial E^\varphi} E^\varphi = 0
\end{equation}
from the coefficient of $(E^\varphi)''$,
\begin{equation}
    \frac{\partial b_2}{\partial E^\varphi} E^{\varphi}
    + \frac{\partial b_2}{\partial K_x} K_x
    + b_2
    = 3 \frac{\partial d_2}{\partial K_\varphi} E^{\varphi }
\end{equation}
from the coefficient of $K_\varphi''$,
\begin{equation}
    \frac{\partial a_2}{\partial E^\varphi} E^{\varphi }
    + \frac{\partial a_2}{\partial K_x} K_x
    + a_2
    = 3 \frac{\partial d_2}{\partial E^x} E^{\varphi}
\end{equation}
from the coefficient of $(E^x)''$,
\begin{equation}
    E^{\varphi} \left( \frac{\partial b_{\varphi \varphi}}{\partial E^\varphi}
    - 2 \frac{\partial c_{\varphi\varphi}}{\partial K_\varphi}
    + 3 \frac{\partial^2 b_2}{(\partial K_\varphi)^2}\right)
    + \frac{\partial b_{\varphi \varphi}}{\partial K_x} K_x
    + b_{\varphi \varphi }
    - 2 \frac{\partial b_2}{\partial K_\varphi}
    = 0
\end{equation}
from the coefficient of $(K_\varphi')^2$,
\begin{equation}
    3 E^{\varphi} \left( \frac{\partial a_{\varphi \varphi}}{\partial E^\varphi}
    - \frac{\partial^2 d_2}{(\partial E^\varphi)^2}\right)
    + a_{\varphi \varphi}
    - \frac{\partial a_{\varphi \varphi}}{\partial K_x} K_x
    = 0
\end{equation}
from the coefficient of $((E^\varphi)')^2$, and
\begin{equation}
    E^{\varphi} \left(-2 \frac{\partial a_{x \varphi}}{\partial E^x}
    + \frac{\partial a_{x x}}{\partial E^\varphi}
    + 3 \frac{\partial^2 d_2}{(\partial E^x)^2}\right)
    + \frac{\partial a_{x x}}{\partial K_x} K_x
    + a_{{\rm lxx}}
    - 2 \frac{\partial a_2}{\partial E^x}
    = 0
\end{equation}
from the coefficient of $((E^x)')^2$.

\section{Restrictions from $\{H,H\}$}
\label{app:HH}

Using $z = K_x / E^\varphi$, the coefficients in (\ref{Gexp}) are
\begin{eqnarray}
    G^\varphi =
    \frac{g^2 E^x}{4 E^\varphi} \left(
    2 A_0 B_{\varphi \varphi}
    - \left(2 z B_{\varphi \varphi} + C_{x \varphi} \right) \frac{\partial A_0}{\partial z}
    + \frac{\partial^2 A_0}{\partial K_\varphi \partial z}
    \right)
    \ ,
\end{eqnarray}
\begin{eqnarray}
    G_x =
    \frac{g^2 E^x}{4 E^\varphi}
    \left(
    A_0 \left( C_{x \varphi}
    - \frac{\partial \ln g}{\partial K_\varphi} \right)
    - \left( 2 A_{x x} + z C_{x \varphi} \right) \frac{\partial A_0}{\partial z}
    - \frac{\partial A_0}{\partial K_\varphi}
    + \frac{\partial^2 A_0}{\partial E^x \partial z}
    \right)
    \ ,
\end{eqnarray}
\begin{eqnarray}
    F_{x x} =
    \frac{g^2 E^x}{4 (E^\varphi)^3}
    \left(
    C_{x \varphi}
    + \frac{\partial \ln g}{\partial K_\varphi}
    + 2 \frac{\partial A_{x x}}{\partial z}
    \right)
    \ ,
\end{eqnarray}
\begin{eqnarray}
    F^\varphi_{x} =
    \frac{g^2 E^x}{4 (E^\varphi)^3} \frac{\partial C_{x \varphi}}{\partial z}
    \ ,
\end{eqnarray}
\begin{eqnarray}
    F^{\varphi \varphi} =
    \frac{g^2 E^x}{4 (E^\varphi)^3} \frac{\partial B_{\varphi \varphi}}{\partial z}
    \ ,
\end{eqnarray}
\begin{eqnarray}
    F^{2\varphi}_x =
    \frac{g^2 E^x}{4 (E^\varphi)^3} \left( 2 B_{\varphi \varphi}
    + \frac{\partial C_{x \varphi}}{\partial z} \right)
    \ ,
\end{eqnarray}

\begin{eqnarray}
    G^{\varphi \varphi x} =
    \frac{g^2 E^x}{4 (E^\varphi)^4} \frac{\partial^2 B_{\varphi \varphi}}{(\partial z)^2}
    \ ,
\end{eqnarray}
\begin{eqnarray}
    G^{x \varphi}_x =
    \frac{g^2 E^x}{4 (E^\varphi)^4} \left( 2 \frac{\partial B_{\varphi \varphi}}{\partial z}
    + \frac{\partial^2 C_{x \varphi}}{(\partial z)^2} \right)
    \ ,
\end{eqnarray}
\begin{eqnarray}
    G_{x x}^x =
    \frac{g^2 E^x}{4 (E^\varphi)^4} \left( \frac{\partial C_{x \varphi}}{\partial z}
    + \frac{\partial^2 A_{x x}}{(\partial z)^2} \right)
    \ ,
\end{eqnarray}
\begin{eqnarray}
    G^{\varphi \varphi}_\varphi =
    - \frac{g^2 E^x}{4 (E^\varphi)^4} \left( 2 \frac{\partial B_{\varphi \varphi}}{\partial z}
    + z \frac{\partial^2 B_{\varphi \varphi}}{(\partial z)^2} \right)
    \ ,
\end{eqnarray}
\begin{eqnarray}
    G^\varphi_{x \varphi} =
    - \frac{g^2 E^x}{4 (E^\varphi)^4} \left( 2  \left(B_{\varphi \varphi} + z \frac{\partial B_{\varphi \varphi}}{\partial z}
    + \frac{\partial C_{x \varphi}}{\partial z} \right)
    + z \frac{\partial^2 C_{x \varphi}}{(\partial z)^2} \right)
    \ ,
\end{eqnarray}
\begin{eqnarray}
    G_{x x \varphi} =
    - \frac{g^2 E^x}{4 (E^\varphi)^4} \left( \frac{\partial \ln g}{\partial K_\varphi}
    + C_{x \varphi}
    + 2 \frac{\partial A_{x x}}{\partial z}
    + z \frac{\partial C_{x \varphi}}{\partial z}
    + z \frac{\partial^2 A_{x x}}{(\partial z)^2}
    \right)
    \ ,
\end{eqnarray}
\begin{eqnarray}
    G^{\varphi \varphi \varphi} =
    \frac{g^2 E^x}{4 (E^\varphi)^3} \left(
    - 2 B_{\varphi \varphi}^2
    - \left( 2 z B_{\varphi \varphi} + C_{x \varphi} \right) \frac{\partial B_{\varphi \varphi}}{\partial z}
    + \frac{\partial^2 B_{\varphi \varphi}}{\partial K_\varphi \partial z}
    \right)
    \ ,
\end{eqnarray}
\begin{eqnarray}
    G^{\varphi \varphi}_x =
    - \frac{g^2 E^x}{4 (E^\varphi)^3} \bigg(
    B_{\varphi \varphi} \left( 3 C_{x \varphi} + \frac{\partial \ln g}{\partial K_\varphi}
    + 2 z \frac{\partial C_{x \varphi}}{\partial z} \right)
    + \left( 2 A_{x x} + z C_{x \varphi} \right) \frac{\partial B_{\varphi \varphi}}{\partial z}
    \nonumber\\
    + C_{x \varphi} \frac{\partial C_{x \varphi}}{\partial z}
    - \frac{\partial B_{\varphi \varphi}}{\partial K_\varphi}
    - \frac{\partial^2 C_{x \varphi}}{\partial K_\varphi \partial z}
    - \frac{\partial^2 B_{\varphi \varphi}}{\partial E^x \partial z}
    \bigg)
    \ ,
\end{eqnarray}
\begin{eqnarray}
    G^\varphi_{x x} =
    - \frac{g^2 E^x}{4 (E^\varphi)^3} \bigg(
    C_{x \varphi}^2
    + 2 A_{x x} \left( B_{\varphi \varphi} + \frac{\partial C_{x \varphi}}{\partial z} \right)
    + C_{x \varphi} \left( \frac{\partial \ln g}{\partial K_\varphi} + \frac{\partial A_{x x}}{\partial z} + z \frac{\partial C_{x \varphi}}{\partial z} \right)
    \nonumber\\
    + 2 z B_{\varphi \varphi} \frac{\partial A_{x x}}{\partial z}
    - \frac{\partial^2 A_{x x}}{\partial K_\varphi \partial z}
    - 2 \frac{\partial B_{\varphi \varphi}}{\partial E^x}
    - \frac{\partial^2 C_{x \varphi}}{\partial E^x \partial z}
    \bigg)
    \ ,
\end{eqnarray}
\begin{eqnarray}
    G_{x x x} =
    - \frac{g^2 E^x}{4 (E^\varphi)^3} \bigg(
    A_{x x} \left( \frac{\partial \ln g}{\partial K_\varphi} + C_{x \varphi} + 2 \frac{\partial A_{x x}}{\partial z} \right)
    + z C_{x \varphi} \frac{\partial A_{x x}}{\partial z}
    + \frac{\partial A_{x x}}{\partial K_\varphi}
    - \frac{\partial C_{x \varphi}}{\partial E^x}
    \frac{\partial A_{x x}}{\partial E^x \partial z}
    \bigg)
    \ ,
\end{eqnarray}

\end{appendix}


\end{document}